\documentclass[10pt,english]{article}

\usepackage[T1]{fontenc}
\usepackage[latin9]{inputenc}
\usepackage[letterpaper]{geometry}
\usepackage{array}
\usepackage{float}
\usepackage{booktabs}
\usepackage{multirow}
\usepackage{amsmath}
\usepackage{amssymb}
\usepackage{graphicx}
\usepackage{setspace}
\usepackage{esint}
\usepackage[numbers]{natbib}
\makeatletter

\providecommand{\tabularnewline}{\\}

\@ifundefined{date}{}{\date{}}
\makeatother

\usepackage{babel}
\begin{document}

\title{\textbf{Relative Resolution:}\\
\textbf{\textcompwordmark{}}\\
\textbf{A Multipole Approximation}\\
\textbf{at Appropriate Distances}\\
\textbf{\textcompwordmark{}}\\
\textcompwordmark{}\\
\vspace{4ex}
}

\author{\textit{Aviel Chaimovich}\\
\textit{Kurt Kremer and Christine Peter}\\
}

\maketitle
\pagebreak{}
\begin{abstract}
Recently, we introduced Relative Resolution as a hybrid formalism
for fluid mixtures \citep{ChaimovichKremer_JCP2015}. The essence
of this approach is that it switches molecular resolution in terms
or relative separation: While nearest neighbors are characterized
by a detailed fine-grained model, other neighbors are characterized
by a simplified coarse-grained model. Once the two models are analytically
connected with each other via energy conservation, Relative Resolution
can capture the structural and thermal behavior of (nonpolar) multi-component
and multi-phase systems across state space. The current work is a
natural continuation of our original communication \citep{ChaimovichKremer_JCP2015}.
Most importantly, we present the comprehensive mathematics of Relative
Resolution, basically casting it as a multipole approximation at appropriate
distances; the current set of equations importantly applies for all
systems (e.g, polar molecules). Besides, we continue examining the
capability of our multiscale approach in molecular simulations, importantly
showing that we can successfully retrieve not just the statics but
also the dynamics of liquid systems. We finally conclude by discussing
how Relative Resolution is the inherent variant of the famous ``cell-multipole''
approach for molecular simulations. 
\end{abstract}
\pagebreak{}
\begin{singlespace}

\section{\textit{Introduction \label{SecA:Introduction}}}
\end{singlespace}

Over the past half of a century, invoking molecular simulations has
become one of the most promising routes for studying soft matter.
This computational approach is splendid for describing systems that
just comprise small spatial and short temporal dimensions (i.e., $<10^{-6}$m
and $<10^{-6}$s, respectively), but it is deficient in describing
systems that also involve large spatial and long temporal dimensions
(i.e., $>10^{-3}$m and $>10^{-3}$s, respectively); resolving such
challenges is especially important for biological processes, whose
particularities usually span orders of magnitude in scale. For overcoming
this dimensionality issue, special attention has been given for enhancing
the computational efficiency of molecular simulations, while ensuring
that the phenomena of interest is still correctly retrieved. One route
involves the intelligent use of statistical mechanics while designing
sophisticated algorithms. For example, various successful strategies
have been developed that focus on improving the computational efficiency
of specific aspects of molecular simulations (e.g., free energies
\citep{Jarzynski_PRL1997,WangLandau_PRL2001,LaioParrinello_PNAS2002,ValssonParrinello_PRL2014},
reaction coefficients \citep{DellagoChandler_JCP1998,AllenWolde_PRL2005,Erp_PRL2007,E0VE0_PRB2002},
etc.). 

Our work revolves around another set of algorithms that has received
much attention in the recent couple of decades: It is commonly called
the multiscale approach, and it generally takes on the in-serial or
in-parallel format. Rather than focusing on the calculation of a specific
feature of a molecular simulation, multiscale algorithms aim at improving
the computational efficiency of the entire system, while ideally capturing
all of its static and dynamic behavior. Importantly, the main signature
of all multiscale simulations is that they involve two systems, one
constructed of detailed Fine-Grained (FG) models with many degrees
of freedom (usually corresponding with atomistic coordinates), and
one constructed of simplified Coarse-Grained (CG) models with few
degrees of freedom (usually corresponding with gravitational centers).
While many multiscale strategies exist that combine quantum-classical
descriptions and discrete-continuum descriptions, the focus of our
publication is on algorithms that combine FG and CG molecular models
based on Newtonian particles. 

The in-serial multiscale methods generally focus on the FG and CG
systems in successive order: In the usual case, they begin by postulating
a particular set of FG models, and after performing a sufficient amount
of molecular simulations of this detailed system, they optimize, based
on a certain criterion, a respective set of CG models, continuing
the remainder of their investigation only with molecular simulations
of this simplified system. While in practice, there are many variants
of this type of multiscale strategies \citep{RuhleAndrienko_JCTC2009,Noid_JCP2013},
each of these can be derived via one of two comprehensive frameworks:
One stems in the relative entropy \citep{Shell_ACP2016}, while the
other aims at ``force-matching'' \citep{NoidAndersen_JCP2008,NoidVoth_JCP2008}.
The former multiscale approach minimizes a functional which measures
the logarithmic ratios between the FG and CG configurational probabilities
\citep{Shell_ACP2016}, and it has been shown that the relative entropy
underlies the uniqueness theorem of Henderson \citep{Henderson_PLA1974},
which is the basis for various strategies that focus on capturing
radial distributions between molecular pairs \citep{GreevyPusztai_MS1988,HG0Stillinger_JCP1993,LyubartsevLaaksonen_PRE1995,ReithMP0_JCC2003}.
The latter multiscale approach minimizes a functional which measures
the squared differences between the FG and CG instantaneous forces,
and it is has been shown that that ``force-matching'' is equivalent
with a generalized version of the Yvon-Born-Green formalism \citep{MullinaxNoid_PRL2009}.
Besides, both of these formalisms retrieve the famous multi-body potential
of Kirkwood \citep{Kirkwood_JCP1935}. 

While all of the in-serial multiscale methods have achieved much success,
they still have several unresolved issues. On a fundamental level,
it is agreed upon that no model can optimally transfer across state
space (e.g., across temperature and density), while also concurrently
representing all structural correlations and thermal properties of
a given system \citep{Louis_JPCM2002,PeterKremer_FD2010}. This technically
means that one must choose a particular aspect of the molecular system
for optimization (e.g., a configurational probability or an instantaneous
force), while having no guarantee of replicating any of its other
behavior, and on top of this, this protocol must be repetitively performed
at each state point of interest. Furthermore on a very practical level,
the most significant challenge with all of these multiscale strategies
is the fact that a very undesirable computational step is always required:
This step involves performing a molecular simulation of a particular
FG model before even using a respective CG model; realize that in
general, if we can construct a molecular simulation of the former,
we do not really need the latter. Finally, we note that because of
the reduction in the degrees of freedom, time-scales naturally change
during multiscale optimization, and thus, kinetic features will inherently
deviate from their true behavior. 

The in-parallel multiscale methods generally contain FG and CG information
simultaneously in a unified system: A single molecular simulation
contains both FG and CG models, with vital aspects described in detail
by the latter, and trivial aspects described with simplicity by the
former. Diverse strategies have been developed that follow this hybrid
computational path, and we categorize them here in four main classes.
The most straightforward approach is the ``group-based'' multiscale
method: Here, essential molecular groups (e.g., solute molecules)
are described via FG interactions, and auxiliary molecular groups
(e.g., solvent molecules) are described via CG interactions \citep{NeriCarloni_PRL2005,ShiVoth_JPCB2006,RzepielaMarrink_PCCP2011,HanSchulten_JCTC2012}.
The next class of strategies can be thought of as the ``time-based''
multiscale method: A given molecular simulation spends some of its
time as the FG system and some of its time as the CG system, with
the results being recorded for the former and discarded for the latter
\citep{HarmandarisKremer_MM2006,SpyriouniMilano_MM2007}. A powerful
variation of this approach actually exchanges among FG and CG replicas
of the molecular simulation, which enhances the overall efficiency
of the computational procedure \citep{LymanZuckerman_PRL2006,ChristenGunsteren_JCP2006}.
In either case, this hybrid approach is obviously convenient for overcoming
various time-scale barriers in a system of interest. 

Perhaps the most successful class of the in-parallel multiscale methods
has been the one which is commonly called Adaptive Resolution. With
much research done on it for over a decade, this approach is a ``one-body
distance-based'' multiscale method: In a single molecular simulation,
adaptive molecules switch their resolution in terms of absolute position
\citep{PraprotnikKremer_JCP2005,Abrams_JCP2005,EnsingParrinello_JCTC2007,PotestioDonadio_PRL2013}.
For example, as molecules move around a system, they adopt a FG model
if near to the origin and a CG model if far from the origin. In practice,
Adaptive Resolution essentially constructs hybrid molecules that embody
both FG and CG models, and in fact, the main difference between all
variations of Adaptive Resolution has been the mixing rule for the
hybrid interaction as a function of absolute position. In the original
publication for Adaptive Resolution, Praprotnik et al.\ introduced
the hybrid interaction as a linear combination of forces, and this
has been the ideal choice for evolving Newtonian trajectories \citep{PraprotnikKremer_JCP2005};
in that same journal issue, Abrams presented an alternative route
that switches between the models in a stochastic manner \citep{Abrams_JCP2005}.
Conversely, Ensing et al.\ formulated the hybrid interaction as a
linear combination of potentials \citep{EnsingParrinello_JCTC2007};
in turn, Potestio et al.\ showed that such a Hamiltonian version
of Adaptive Resolution naturally corresponds with many aspects in
statistical mechanics \citep{PotestioDonadio_PRL2013}. On a practical
level, Adaptive Resolution is especially useful if a given system
has a specific region of special interest (e.g., a protein at the
origin, with everything else being water). As such, Adaptive Resolution
has been already applied in some biological scenarios, having quite
successful results \citep{ZavadlavPraprotnik_JCP2014,FogartyKremer_JCP2015}. 

Very recently, we developed another type of multiscale methods, termed
Relative Resolution (RelRes) \citep{ChaimovichKremer_JCP2015}. Inspired
by Adaptive Resolution, this approach is a ``two-body distance-based''
multiscale method: In a single molecular simulation, hybrid molecules
switch their resolution in terms of relative separation; in particular,
molecules that are near neighbors (i.e., their pairwise distance is
small) interact via their FG models, and molecules that are far neighbors
(i.e., their pairwise distance is large) interact via their CG models
\citep{ChaimovichKremer_JCP2015}. Importantly, RelRes is the sole
class of multiscale simulations which describes all molecules, for
all times, with both FG and CG models; the resolution of a given molecule
is always relative of its observer. While other strategies, reminiscent
of RelRes, have been also developed \citep{IzvekovVoth_JCTC2009,ShenHu_JCTC2014},
our formalism is unique in that it mathematically finds a natural
connection between the FG and CG potentials. While Ref.\ \citep{IzvekovVoth_JCTC2009}
numerically parametrizes between the two models via ``force-matching'',
Ref.\ \citep{ShenHu_JCTC2014} does not make a clear connection between
the two models. Conversely, we developed an analytical parameterization
between the FG and CG models, which is just based on energy conservation
\citep{ChaimovichKremer_JCP2015}. Unlike those other strategies \citep{IzvekovVoth_JCTC2009,ShenHu_JCTC2014},
we were consequently able of correctly retrieving across state space,
the structural and thermal behavior of several nonpolar mixtures \citep{ChaimovichKremer_JCP2015}.
Besides, we showed that our hybrid approach can be considered as a
generalized extension of established theories for uniform liquids,
which assume a mean field for interactions beyond a certain distance
\citep{Frisch_ACP1964,Widom_Science1967,WeeksAndersen_JCP1971,IngebrigtsenDyre_PRX2012}. 

In this current work, we build on our original communication of RelRes
\citep{ChaimovichKremer_JCP2015}. Foremost, we mathematically present
the comprehensive framework for RelRes. Above all by performing a
multipole expansion for an arbitrary potential of inverse powers,
we show that its zero-order term, which is sufficient for nonpolar
systems, is identical with the succinct expression we presented in
our initial publication; in consideration of polar systems, the current
work also presents the first-order and second-order terms of the Taylor
series, both of which involve molecular orientation. By use of this
multipole approximation, we introduce RelRes as a Hamiltonian that
switches between the FG and CG models at an appropriate distance.
Building on the computational results of our original work, we consequently
continue examining RelRes with molecular simulations of nonpolar systems.
Interestingly, we find that the ideal switching distance between the
FG and CG models is roughly equivalent with the location at which
a particular orientational function essentially decorrelates. Besides,
we also demonstrate that RelRes can correctly describe not just static
behavior but also dynamic behavior. Together with our previous publication
\citep{ChaimovichKremer_JCP2015}, we suggest that RelRes can be a
powerful multiscale tool for efficiently studying soft matter via
molecular simulations. 

\pagebreak{}
\begin{singlespace}

\section{\textit{Theoretical Foundation \label{SecA:Theoretical}}}
\end{singlespace}

Contrary to our original publication \citep{ChaimovichKremer_JCP2015},
we present here our hybrid formalism in a reversed order. In the previous
work, we started by defining RelRes, and we ended by parameterizing
between the FG and CG potentials. In the current work, we start by
parameterizing between the FG and CG potentials, and we end by defining
RelRes. Regardless, the mathematical notation is identical between
our two publications. Note also that the current framework is the
comprehensive one, applying for all molecular systems. The original
publication is the terse version, which is only applicable for nonpolar
scenarios. 

\vspace{4ex}

\subsection{\textit{Defining our System \label{SecB:Theoretical1} }}

\hspace{2em}Consider that we are performing a molecular simulation
of a pair of molecules in vacuum; a particular configuration of this
trivial system is depicted in Fig.\ \ref{Fig:MPIL}. We label the
gravitational centers by the Latin indices $i$ and $j$, and we label
the atomistic coordinates by the Greek indices $\mu$ and $\nu$.
In our notation throughout, we denote $n_{i}$ or $n_{j}$ as the
number of sites on a molecule $i$ or $j$, respectively. The relative
separation between the centers is $\overrightarrow{r}_{ij}$, and
the relative separation between the coordinates is$\overrightarrow{r}_{\mu_{i}\nu_{j}}$.
Furthermore, the distance between a certain atomistic coordinate and
a respective gravitational center is $\overrightarrow{\Delta}_{\mu_{i}}$
or $\overrightarrow{\Delta}_{\nu_{j}}$. By Fig.\ \ref{Fig:MPIL},
it is clear that these distances are related as follows, 
\begin{equation}
\overrightarrow{\Delta}_{\mu_{i}\nu_{j}}=\overrightarrow{r}_{\mu_{i}\nu_{j}}-\overrightarrow{r}_{ij}\label{Eq:VctorDstnc}
\end{equation}
and we defined in this expression the following variable:

\begin{equation}
\overrightarrow{\Delta}_{\mu_{i}\nu_{j}}=\overrightarrow{\Delta}_{\nu_{j}}-\overrightarrow{\Delta}_{\mu_{i}}\label{Eq:Delta}
\end{equation}
Note that we introduce here all distances in vector form, and their
scalar definitions naturally follow. We also define dimensionless
variables that will compact all of our ensuing mathematics:

\begin{equation}
\overrightarrow{\xi}_{\mu_{i}\nu_{j}}=\frac{\overrightarrow{\Delta}_{\mu_{i}\nu_{j}}}{r_{ij}}\label{Eq:DmnlsXi}
\end{equation}
\begin{equation}
\cos\theta_{\mu_{i}\nu_{j}}=\frac{\overrightarrow{r}_{ij}\cdot\overrightarrow{\Delta}_{\mu_{i}\nu_{j}}}{r_{ij}\Delta_{\mu_{i}\nu_{j}}}\label{Eq:DmnlsTh}
\end{equation}
With these definitions, together with some rearrangement, we attain
this useful relation which basically stems in the dot product of Eq.\ \ref{Eq:VctorDstnc}
with itself:

\begin{equation}
\frac{r_{\mu_{i}\nu_{j}}}{r_{ij}}=\sqrt{1+2\xi_{\mu_{i}\nu_{j}}\cos\theta_{\mu_{i}\nu_{j}}+\xi_{\mu_{i}\nu_{j}}^{2}}\label{Eq:SclarDstncA}
\end{equation}
Finally for clarity, we must make a subtle distinction with a variable,
$\Delta r_{\mu_{i}\nu_{j}}$, that appears in our original communication,
but that we do not use here: $\Delta r_{\mu_{i}\nu_{j}}=r_{\mu_{i}\nu_{j}}-r_{ij}\neq\Delta{}_{\mu_{i}\nu_{j}}$. 

We now focus on the energetics of this rudimentary system. The governing
function here is defined as follows: 

\begin{equation}
u_{ij}\left(\overrightarrow{r}_{ij};\left\{ \overrightarrow{\Delta}_{\mu_{i}}\right\} ,\left\{ \overrightarrow{\Delta}_{\nu_{j}}\right\} \right)=\sum_{\mu_{i}\nu_{j}}{\displaystyle u_{\mu_{i}\nu_{j}}\left(r_{\mu_{i}\nu_{j}}\right)}\label{Eq:EnergyBase}
\end{equation}
Here, $u_{\mu_{i}\nu_{j}}$ is the intrinsic potential between atomistic
coordinates $\mu_{i}$ and $\nu_{j}$, and $u_{ij}$ is the resultant
potential between gravitational centers $i$ and $j$; the latter
is obtained by the respective summation of the former. Notice that
we are allowing for absolute nonuniformity in our mixtures considering
the indices on the potentials, $u_{\mu_{i}\nu_{j}}$ and $u_{ij}$.
Importantly, we assume that the former is isotropic, being exclusively
a function of the scalar $r_{\mu_{i}\nu_{j}}$. Conversely, we expect
that the latter is geometric, being chiefly a function of the vector
$\overrightarrow{r}_{ij}$, as well as the two sets of intramolecular
distances, $\left\{ \overrightarrow{\Delta}_{\mu_{i}}\right\} $ and
$\left\{ \overrightarrow{\Delta}_{\nu_{j}}\right\} $. Importantly,
we ignore intramolecular energetics during most of our derivation:
These are accounted for once we present the entire Hamiltonian of
RelRes. 

One of the focal assumptions in our analysis is that the basis function
of the potential can be cast as an inverse power of the isotropic
distance: 

\begin{equation}
u_{\mu_{i}\nu_{j}}\left(r\right)=\frac{c_{\mu_{i}\nu_{j}}}{r^{m}}\label{Eq:InversePower}
\end{equation}
Obviously, $m$ is the power associated with this inverse law. In
most scenarios, like with the Coulomb potential ($m=1$), as well
as with the Lennard-Jones (LJ) potential ($m=12$ and $m=6$), the
inverse law is a natural choice. The proportionality coefficient $c_{\mu_{i}\nu_{j}}$
corresponds with the unique parameters involved in the interaction
between sites $\mu_{i}$ and $\nu_{j}$, and it is usually known empirically
for use in molecular simulations. We will later make an important
assumption of geometric mixing for $c_{\mu_{i}\nu_{j}}$ (i.e., we
will express it as a product of two separate parameters).

In the remainder of the derivation, our goal is to reformulate $u_{ij}$
of Eq.\ \ref{Eq:EnergyBase} in terms of $r_{ij}$, together with
variations of $\overrightarrow{\xi}_{\mu_{i}\nu_{j}}$ and $\cos\theta_{\mu_{i}\nu_{j}}$,
as defined by Eqs.\ \ref{Eq:DmnlsXi} and \ref{Eq:DmnlsTh}. Such
a reformulation holds a promise for the efficient calculation of the
energy of the pair in Fig.\ \ref{Fig:MPIL}; obviously, computation
of a single distance, $r_{ij}$, is preferable over the computation
of $n_{i}n_{j}$ distances, $r_{\mu_{i}\nu_{j}}$, which the summation
of Eq.\ \ref{Eq:EnergyBase} requires. Of course, such a reformulation
may involve approximations, and thus, we must make valid assumptions
for $\overrightarrow{\xi}_{\mu_{i}\nu_{j}}$ and $\cos\theta_{\mu_{i}\nu_{j}}$
so that we maintain the correct energy for this pair. 

\subsection{\textit{Introducing the Multipole Expansion }\textit{\normalsize{}\label{SecB:Theoretical2}}}

\hspace{2em}We substitute Eq.\ \ref{Eq:InversePower} in Eq.\ \ref{Eq:EnergyBase},
which yields the following: 

\begin{equation}
u_{ij}\left(\overrightarrow{r}_{ij};\overrightarrow{\xi}_{\mu_{i}\nu_{j}},\cos\theta_{\mu_{i}\nu_{j}}\right)=\sum_{\mu_{i}\nu_{j}}\frac{c_{\mu_{i}\nu_{j}}}{r_{\mu_{i}\nu_{j}}^{m}}\label{Eq:EnergyA}
\end{equation}
For clarity throughout most of the ensuing analysis, $\overrightarrow{\xi}_{\mu_{i}\nu_{j}}$
and $\cos\theta_{\mu_{i}\nu_{j}}$ are omitted yet implied in the
functionalities involved in Eq.\ \ref{Eq:EnergyA}; the same also
goes for the bar on $\overrightarrow{r}_{ij}$. Consider now the radial
distance which appears here together with the power of $m$; we correspondingly
exponentiate Eq.\ \ref{Eq:SclarDstncA}, and we present the ensuing
reciprocal: 

\begin{equation}
\left(\frac{r_{ij}}{r_{\mu_{i}\nu_{j}}}\right)^{m}=\left(1+2\xi_{\mu_{i}\nu_{j}}\cos\theta_{\mu_{i}\nu_{j}}+\xi_{\mu_{i}\nu_{j}}^{2}\right)^{-m/2}\label{Eq:SclarDstncB}
\end{equation}
We can use this expression to substitute the computationally inexpensive
$r_{ij}$ for the computationally expensive $r_{\mu_{i}\nu_{j}}$
in Eq.\ \ref{Eq:EnergyA}. We consequently attain the following expression: 

\begin{equation}
u_{ij}\left(r_{ij}\right)=\frac{1}{r_{ij}^{m}}\sum_{\mu_{i}\nu_{j}}\left[c_{\mu_{i}\nu_{j}}\cdot\left(1+2\xi_{\mu_{i}\nu_{j}}\cos\theta_{\mu_{i}\nu_{j}}+\xi_{\mu_{i}\nu_{j}}^{2}\right)^{-m/2}\right]\label{Eq:EnergyB}
\end{equation}

Analogous with the usual multipole expansion, we now reformulate,
via a Taylor series, the cumbersome expression which appears above, 

\begin{equation}
\left(1+2\xi\cos\theta+\xi^{2}\right)^{-m/2}=\sum_{\aleph}\left[\Lambda_{m,\aleph}\left(\cos\theta\right)\cdot\xi^{\aleph}\right]\label{Eq:TaylorSeries}
\end{equation}
with $\Lambda_{m,\aleph}$ basically defined by the appropriate derivative
in terms of the index $\aleph$: 

\begin{equation}
\Lambda_{m,\aleph}\left(\cos\theta\right)=\frac{1}{\aleph!}\left[\frac{\partial^{\aleph}}{\partial\xi^{\aleph}}\left(1+2\xi\cos\theta+\xi^{2}\right)^{-m/2}\right]_{\xi=0}\label{Eq:TaylorTerm}
\end{equation}
Clearly, $\Lambda_{m,\aleph}$ is only a function of $\cos\theta_{\mu_{i}\nu_{j}}$,
with the functionality of $\xi_{\mu_{i}\nu_{j}}$ appearing as a power
in Eq.\ \ref{Eq:TaylorSeries}. Realize that such a Taylor expansion
is justified by the fact that usually, $\xi_{\mu_{i}\nu_{j}}<1$ (i.e.,
$\Delta_{\mu_{i}\nu_{j}}<r_{ij}$), since in most cases, bonds are
negligible in magnitude compared with intermolecular distances. Notice
also that the exponent $m$ of the inverse law is just a parameter
in $\Lambda_{m,\aleph}$. Besides, if $m=1$ (i.e., the Coulomb potential),
Eq.\ \ref{Eq:TaylorSeries} simply becomes the common definition
of the Legendre polynomials. 

Substituting Eq.\ \ref{Eq:TaylorSeries} in Eq.\ \ref{Eq:EnergyB}
yields the following power series for our energy function:

\begin{equation}
u_{ij}\left(r_{ij}\right)=\frac{1}{r_{ij}^{m}}\sum_{\aleph}\left[\sum_{\mu_{i}\nu_{j}}\left[c_{\mu_{i}\nu_{j}}\Lambda_{m,\aleph}\left(\cos\theta_{\mu_{i}\nu_{j}}\right)\xi_{\mu_{i}\nu_{j}}^{\aleph}\right]\right]\label{Eq:EnergyC}
\end{equation}
and it is particularly convenient to cast the above as follows, 

\begin{equation}
u_{ij}\left(r_{ij}\right)=\frac{1}{r_{ij}^{m}}\sum_{\aleph}c_{ij,\aleph}\left(\xi_{\mu_{i}\nu_{j}},\cos\theta_{\mu_{i}\nu_{j}}\right)\label{Eq:EnergyD}
\end{equation}
in which we obviously defined the following:

\begin{equation}
c_{ij,\aleph}\left(\xi_{\mu_{i}\nu_{j}},\cos\theta_{\mu_{i}\nu_{j}}\right)=\sum_{\mu_{i}\nu_{j}}\left[c_{\mu_{i}\nu_{j}}\Lambda_{m,\aleph}\left(\cos\theta_{\mu_{i}\nu_{j}}\right)\xi_{\mu_{i}\nu_{j}}^{\aleph}\right]\label{Eq:CoefficientTerm}
\end{equation}
One may think of $c_{ij,\aleph}$ as an effective coefficient for
a given configuration of Fig.\ \ref{Fig:MPIL}; analogous with $u_{ij}$,
$\xi_{\mu_{i}\nu_{j}}$ and $\cos\theta_{\mu_{i}\nu_{j}}$ are consequently
omitted yet implied in the functionality of $c_{ij,\aleph}$ throughout
most of our derivation. In any case, notice that these expressions
are especially useful for computational purposes: By summing over
all sites $\mu_{i}$ and $\nu_{j}$, one can separately calculate
the energetic coefficient of any $\aleph$ term by Eq.\ \ref{Eq:CoefficientTerm},
and by including as many $\aleph$ terms in Eq.\ \ref{Eq:EnergyD}
as desired, one can have a valid approximation for the overall interaction
between the molecular pair of Fig.\ \ref{Fig:MPIL}. Of particular
importance is the leading (nonvanishing) $\aleph$ term: For making
a connection with the terminology of the original work, we term it
as the Molecular Pair in the Infinite Limit (MPIL). 

\subsection{\textit{Evaluating the Multipole Terms }\textit{\normalsize{}\label{SecB:Theoretical3}}}

\hspace{2em}By not even performing any differentiation, the $\aleph=0$
term is readily determined by Eq.\ \ref{Eq:TaylorTerm} as the following: 

\begin{equation}
\Lambda_{m,0}=1\label{Eq:TaylorTerm0}
\end{equation}
Conveniently, this is always unity, regardless of the value of $m$.
Plugging this in Eq.\ \ref{Eq:CoefficientTerm}, we attain this expression:

\begin{equation}
c_{ij,0}=\sum_{\mu_{i}\nu_{j}}c_{\mu_{i}\nu_{j}}\label{Eq:CoefficientTerm0}
\end{equation}
Notice here that this is not dependent at all on the configuration
of the molecular pair: This means $c_{ij,0}$ is a constant that can
be calculated before any molecular simulation is performed. If this
is nonzero, we can approximate our energy in Eq.\ \ref{Eq:EnergyD}: 

\begin{equation}
u_{ij}\left(r_{ij}\right)\approx\frac{c_{ij,0}}{r_{ij}^{m}}\label{Eq:Energy0}
\end{equation}

\[
u_{ij}\left(r_{ij}\right)\approx\frac{\sum_{\mu_{i}\nu_{j}}c_{\mu_{i}\nu_{j}}}{r_{ij}^{m}}
\]
In the later part of this expression, we obviously invoked Eq.\ \ref{Eq:CoefficientTerm0}.
Notice the similarity between Eqs.\ \ref{Eq:EnergyA} and \ref{Eq:Energy0}:
We start with an $m$ inverse law in the former, and we end with an
$m$ inverse law in the latter. In Eq.\ \ref{Eq:EnergyA}, we sum
over the many distances between the atomistic coordinates, with their
intrinsic interaction coefficients, and in Eq.\ \ref{Eq:Energy0},
we take in the single distance between the gravitational centers,
with its resultant interaction coefficient. This is in effect the
monopole term; once we introduce another assumption, we will show
that this is specifically equivalent with the monopole-monopole interaction. 

This is in essence the sole term that we derived in our previous work.
We simply attained it by vanishing all bond lengths (i.e., $\Delta_{\mu_{i}\nu_{j}}=0$)
in Fig.\ \ref{Fig:MPIL}; this is basically equivalent to setting
all $\xi_{\mu_{i}\nu_{j}}=0$ in Eq.\ \ref{Eq:EnergyB}, and in turn,
it also eliminates the corresponding functionality of all $\cos\theta_{\mu_{i}\nu_{j}}$
in our energy expression. The only discrepancy of Eq.\ \ref{Eq:Energy0}
with the corresponding one in our original work is that in this one,
we have the assumption of the inverse power for the interaction. We
can remove this assumption temporarily by invoking Eq.\ \ref{Eq:InversePower}
for Eq.\ \ref{Eq:Energy0}, and we in turn get the exact expression
which we termed as the MPIL in our original publication:

\begin{equation}
u_{ij}\left(r_{ij}\right)\approx\sum_{\mu_{i}\nu_{j}}u_{\mu_{i}\nu_{j}}\left(r_{ij}\right)\label{Eq:Energy00}
\end{equation}
Realize that that this is a special case of our current definition
of the MPIL: As mentioned before, the general definition of our MPIL
is the leading (nonvanishing) $\aleph$ term in Eq.\ \ref{Eq:EnergyC}. 

Importantly in our original work, we showed that if Eq.\ \ref{Eq:Energy0}
is implemented with RelRes, it can successfully describe nonpolar
mixtures that are based on the LJ potential. Theoretically, this approximation
shall also work with the Coulomb potential for ionic molecules that
have a net charge. In fact for RelRes, we expect that Eq.\ \ref{Eq:Energy0}
shall be adequate for any $m$, given that this monopole term does
not vanish. Nevertheless, once the above summation for a certain molecular
pair contains both positive and negative values for $c_{\mu_{i}\nu_{j}}$,
there may be some issues; specifically, if all $c_{\mu_{i}\nu_{j}}$
in the summation cancel each other, Eq.\ \ref{Eq:Energy0} is strictly
zero. The most obvious such case is the zwitterionic scenario, in
which the molecules have Coulombic charges, yet they are overall neutral
(e.g., water). It is obvious that the interaction between such polar
pairs is finite, and consequently, one must go beyond the monopole
term of Eq.\ \ref{Eq:Energy0} in order to describe the corresponding
energy, at least approximately. This is in fact the purpose of the
ensuing mathematics. Note that the equations below may be also useful
as correction terms for the interaction between the molecular pair,
even if their monopole term is nonzero. 

Obviously, all the terms of the Taylor series can be evaluated by
successive differentiation as presented in Eq.\ \ref{Eq:TaylorTerm}.
Here is the first-order term: 

\[
\Lambda_{m,1}\left(\cos\theta\right)=\left[m\left(\cos\theta+\xi\right)\left(1+2\xi\cos\theta+\xi^{2}\right)^{-\left(m+2\right)/2}\right]_{\xi=0}
\]

\begin{equation}
\Lambda_{m,1}\left(\cos\theta\right)=m\cos\theta\label{Eq:TaylorTerm1}
\end{equation}
Consequently, the $\aleph=1$ term of Eq.\ \ref{Eq:CoefficientTerm}
is the following,

\begin{equation}
c_{ij,1}\left(\xi_{\mu_{i}\nu_{j}},\cos\theta_{\mu_{i}\nu_{j}}\right)=m\sum_{\mu_{i}\nu_{j}}\left[c_{\mu_{i}\nu_{j}}\cos\theta_{\mu_{i}\nu_{j}}\xi_{\mu_{i}\nu_{j}}\right]\label{Eq:CoefficientTerm1}
\end{equation}
Here is the second-order term: 

\[
\Lambda_{m,2}\left(\cos\theta\right)=\frac{1}{2}\left[m\left(m+2\right)\left(\cos\theta+\xi\right)^{2}\left(1+2\xi\cos\theta+\xi^{2}\right)^{-\left(m+4\right)/2}-m\left(1+2\xi\cos\theta+\xi^{2}\right)^{-\left(m+2\right)/2}\right]_{\xi=0}
\]

\begin{equation}
\Lambda_{m,2}\left(\cos\theta\right)=\frac{m}{2}\left(\left(m+2\right)\cos^{2}\theta-1\right)\label{Eq:TaylorTerm2}
\end{equation}
Consequently, the $\aleph=2$ term of Eq.\ \ref{Eq:CoefficientTerm}
is the following,

\begin{equation}
c_{ij,2^{+}}\left(\xi_{\mu_{i}\nu_{j}},\cos\theta_{\mu_{i}\nu_{j}}\right)=\frac{m}{2}\sum_{\mu_{i}\nu_{j}}\left[c_{\mu_{i}\nu_{j}}\cos^{2}\theta_{\mu_{i}\nu_{j}}\xi_{\mu_{i}\nu_{j}}^{2}\right]\label{Eq:CoefficientTerm2+}
\end{equation}

\begin{equation}
c_{ij,2^{-}}\left(\xi_{\mu_{i}\nu_{j}},\cos\theta_{\mu_{i}\nu_{j}}\right)=\frac{m}{2}\sum_{\mu_{i}\nu_{j}}\left[c_{\mu_{i}\nu_{j}}\xi_{\mu_{i}\nu_{j}}^{2}\right]\label{Eq:CoefficientTerm2-}
\end{equation}
Here, we introduced this definition: 

\begin{equation}
c_{ij,2}\left(\xi_{\mu_{i}\nu_{j}},\cos\theta_{\mu_{i}\nu_{j}}\right)=\left(m+2\right)c_{ij,2^{+}}-c_{ij,2^{-}}\label{Eq:CoefficientTerm2}
\end{equation}
We also emphasize here that if $m=1$ (e.g., the Coulombic scenario),
$\Lambda_{m,1}$ and $\Lambda_{m,2}$ become the familiar first and
second Legendre polynomials, respectively. 

Of course, these $c_{ij,\aleph}$ can be readily used for evaluating
the energy function of Eq.\ \ref{Eq:EnergyD}: They may serve as
correction terms for Eq.\ \ref{Eq:Energy0}, and if $c_{ij,1}$ or
$c_{ij,2}$ is actually the leading term of this multipole expansion,
it can even become the sole approximation for the energy (i.e., the
MPIL). Notice that unlike $c_{ij,0}$, these current expressions are
functions of $\cos\theta_{\mu_{i}\nu_{j}}$, and well as $\xi_{\mu_{i}\nu_{j}}$.
Thus, while Eq.\ \ref{Eq:Energy0} is isotropic, involving just the
scalar $r_{ij}$ in its functionality, the energy which corresponds
with other $c_{ij,\aleph}$ is geometric, involving also the vector
$\overrightarrow{\Delta}_{\mu_{i}\nu_{j}}$ in its functionality.
Thus for $\aleph\neq0$, the coefficient must be computed at each
step of a molecular simulation. As such, going beyond the monopole
term obviously requires significant computational cost, and it is
only recommended if it is necessary (i.e., for polar systems in which
these are the leading terms in the Taylor series). 

\subsection{\textit{Assuming a Proportionality Coefficient }\textit{\normalsize{}\label{SecB:Theoretical4}}}

\hspace{2em}Let us now discuss the proportionality coefficient $c_{\mu_{i}\nu_{j}}$
of the inverse law of Eq.\ \ref{Eq:InversePower}. In most cases,
such a parameter is empirically known for use in molecular simulations.
In fact, it is usually expressed in terms of some mixing rules between
two separate coefficients $\mathfrak{c}_{\mu_{i}}$ and $\mathfrak{c}_{\nu_{j}}$
of sites $\mu_{i}$ and $\nu_{j}$, respectively. For example with
the Coulomb potential, $c_{\mu_{i}\nu_{j}}$ is fundamentally the
product of partial charges; an analogous product is also sometimes
used for the LJ potential. For the remainder of our work, we assume
that the following equation holds: 

\begin{equation}
c_{\mu_{i}\nu_{j}}=\mathfrak{c}_{\mu_{i}}\mathfrak{c}_{\nu_{j}}\label{Eq:Coefficient}
\end{equation}
For completeness, this means that Eq.\ \ref{Eq:EnergyA} (i.e., the
initial energy function in our derivation that introduced the inverse
power), becomes the following: 

\begin{equation}
u_{ij}\left(\overrightarrow{r}_{ij};\overrightarrow{\xi}_{\mu_{i}\nu_{j}},\cos\theta_{\mu_{i}\nu_{j}}\right)=\sum_{\mu_{i}\nu_{j}}\frac{\mathfrak{c}_{\mu_{i}}\mathfrak{c}_{\nu_{j}}}{r_{\mu_{i}\nu_{j}}^{m}}\label{Eq:EnergyE}
\end{equation}
In any case, Eq.\ \ref{Eq:Coefficient} substantially facilitates
our computational algorithm for RelRes, especially since a variant
of $c_{ij,\aleph}$, for any $\aleph$, becomes a constant during
a molecular simulation. 

Let us begin with the zero-order term. Substituting Eq.\ \ref{Eq:Coefficient}
in Eq.\ \ref{Eq:CoefficientTerm0}, we attain the following equation 

\begin{equation}
c_{ij,0}=\mathfrak{c}_{i,0}\mathfrak{c}_{j,0}\label{Eq:CoefficientTerm0_}
\end{equation}
in which we have introduced the definition of the monopole $\mathfrak{c}_{i,0}$
or $\mathfrak{c}_{j,0}$, for each molecule $i$ or $j$, respectively: 

\begin{equation}
\begin{array}{c}
\mathfrak{c}_{i,0}=\sum_{\mu_{i}}\mathfrak{c}_{\mu_{i}}\\
\mathfrak{c}_{j,0}=\sum_{\nu_{j}}\mathfrak{c}_{\nu_{j}}
\end{array}\label{Eq:Monopole}
\end{equation}
If both of these monopoles are nonzero, the MPIL is the following
according to Eq.\ \ref{Eq:EnergyD}:

\begin{equation}
u_{ij}\left(r_{ij}\right)\approx\frac{\mathfrak{c}_{i,0}\mathfrak{c}_{j,0}}{r_{ij}^{m}}\label{Eq:Energy0_}
\end{equation}

\[
u_{ij}\left(r_{ij}\right)\approx\frac{\left(\sum_{\mu_{i}}\mathfrak{c}_{\mu_{i}}\right)\left(\sum_{\nu_{j}}\mathfrak{c}_{\nu_{j}}\right)}{r_{ij}^{m}}
\]
Analogous with our comparison of Eqs.\ \ref{Eq:EnergyA} and \ref{Eq:Energy0}
above, notice the similarity between Eqs.\ \ref{Eq:EnergyE} and
\ref{Eq:Energy0_}: Performing the approximation, we start with several
inverse coefficients $\mathfrak{c}_{\mu_{i}}\mathfrak{c}_{\nu_{j}}$,
and we end with one inverse coefficient $\mathfrak{c}_{i,0}\mathfrak{c}_{j,0}$. 

We now consider the benefit of the product assumption of Eq.\ \ref{Eq:Coefficient}.
You may notice that Eq.\ \ref{Eq:Energy0_} may be slightly more
computationally efficient than Eq.\ \ref{Eq:Energy0}. The former
involves two summations over $n_{i}$ and $n_{j}$ parameters, while
the latter involves one summation over $n_{i}n_{j}$ parameters. Regardless,
all of these summations can be performed before a molecular simulation,
and thus, they have negligible computational cost. The computational
superiority introduced via Eq.\ \ref{Eq:Coefficient} becomes very
apparent once we move to the first-order and second-order terms of
the multipole expansion. 

We may now apply the assumption of Eq.\ \ref{Eq:Coefficient} for
evaluating any $c_{ij,\aleph}$, as presented in Eq.\ \ref{Eq:CoefficientTerm},
after the appropriate differentiation of Eq.\ \ref{Eq:TaylorTerm}.
As an important part of the ensuing mathematics, we introduce the
familiar definition for the dipole $\overrightarrow{\mathfrak{p}}_{i,1}$
or $\overrightarrow{\mathfrak{p}}_{j,1}$, of each molecule $i$ or
$j$, respectively: 

\begin{equation}
\begin{array}{c}
\overrightarrow{\mathfrak{p}}_{i,1}=\sum_{\mu_{i}}\left[\mathfrak{c}_{\mu_{i}}\overrightarrow{\Delta}_{\mu_{i}}\right]\\
\overrightarrow{\mathfrak{p}}_{j,1}=\sum_{\nu_{j}}\left[\mathfrak{c}_{\nu_{j}}\overrightarrow{\Delta}_{\nu_{j}}\right]
\end{array}\label{Eq:Dipole}
\end{equation}
Moreover, we introduce components of the familiar definition for the
quadrupole $\left(\overrightarrow{\overrightarrow{\mathfrak{q}}}_{i,2},\mathfrak{q}_{i,0}\right)$
or $\left(\overrightarrow{\overrightarrow{\mathfrak{q}}}_{j,2},\mathfrak{q}_{j,0}\right)$,
of each molecule $i$ or $j$, respectively: 

\begin{equation}
\begin{array}{c}
\overrightarrow{\overrightarrow{\mathfrak{q}}}_{i,2}=\sum_{\mu_{i}}\left[\mathfrak{c}_{\mu_{i}}\overrightarrow{\Delta}_{\mu_{i}}\overrightarrow{\Delta}_{\mu_{i}}\right]\\
\overrightarrow{\overrightarrow{\mathfrak{q}}}_{j,2}=\sum_{\nu_{j}}\left[\mathfrak{c}_{\nu_{j}}\overrightarrow{\Delta}_{\nu_{j}}\overrightarrow{\Delta}_{\nu_{j}}\right]
\end{array}\label{Eq:Quadrupole+}
\end{equation}

\begin{equation}
\begin{array}{c}
\mathfrak{q}_{i,0}=\sum_{\mu_{i}}\left[\mathfrak{c}_{\mu_{i}}\Delta_{\mu_{i}}^{2}\right]\\
\mathfrak{q}_{j,0}=\sum_{\nu_{j}}\left[\mathfrak{c}_{\nu_{j}}\Delta_{\nu_{j}}^{2}\right]
\end{array}\label{Eq:Quadrupole-}
\end{equation}
By casting a linear combination of these, $\left(m+2\right)\overrightarrow{\overrightarrow{\mathfrak{q}}}_{i,2}-\mathfrak{q}_{i,0}\overrightarrow{\overrightarrow{\delta}}$
or $\left(m+2\right)\overrightarrow{\overrightarrow{\mathfrak{q}}}_{j,2}-\mathfrak{q}_{j,0}\overrightarrow{\overrightarrow{\delta}}$,
we retrieve the conventional definition of the quadruple for each
molecule $i$ or $j$, respectively; this is most apparent for $m=1$
(e.g, the Coulombic scenario). As in the usual case, we are permitting
here for generalized tensor algebra (e.g., $\overrightarrow{\overrightarrow{\mathcal{I}}}$
is the identity tensor); the amount of bars on these parameters, together
with their last index for extra emphasis, corresponds with the order
of the tensor. For compactness, we introduce here a compact notation
for the set of dipoles (i.e., $\overrightarrow{\mathfrak{p}}_{ij,1}=\left\{ \overrightarrow{\mathfrak{p}}_{i,1},\overrightarrow{\mathfrak{p}}_{j,1}\right\} $),
as well as for the set of quadrupoles (i.e., $\overrightarrow{\overrightarrow{\mathfrak{q}}}_{ij,2}=\left\{ \overrightarrow{\overrightarrow{\mathfrak{q}}}_{i,2},\overrightarrow{\overrightarrow{\mathfrak{q}}}_{j,2}\right\} $
and $\mathfrak{q}_{ij,0}=\left\{ \mathfrak{q}_{i,0},\mathfrak{q}_{j,0}\right\} $).
As we all know, the elegance of each of these parameters is that it
is only a function of the topology of each molecule, and it does not
at all depend on the overall configuration of the molecular pair in
Fig.\ \ref{Fig:MPIL}. 

By invoking these, we specifically show in the Appendix for $\aleph=1$
and $\aleph=2$ how the following expressions for $c_{ij,\aleph}$
can be derived. Here is the first-order term: 

\begin{equation}
c_{ij,1}\left(\overrightarrow{\mathfrak{p}}_{ij,1}\right)=\frac{m}{r_{ij}}\left[\mathfrak{c}_{i,0}\left(\hat{r}_{ij}\cdot\overrightarrow{\mathfrak{p}}_{j,1}\right)-\mathfrak{c}_{j,0}\left(\hat{r}_{ij}\cdot\overrightarrow{\mathfrak{p}}_{i,1}\right)\right]\label{Eq:CoefficientTerm1_}
\end{equation}
Here is the second-order term:

\begin{equation}
c_{ij,2^{+}}\left(\overrightarrow{\mathfrak{p}}_{ij,1},\overrightarrow{\overrightarrow{\mathfrak{q}}}_{ij,2}\right)=\frac{m}{r_{ij}^{2}}\left[-\left(\hat{r}_{ij}\cdot\overrightarrow{\mathfrak{p}}_{i,1}\right)\left(\hat{r}_{ij}\cdot\overrightarrow{\mathfrak{p}}_{j,1}\right)+\frac{1}{2}\mathfrak{c}_{i,0}\left(\hat{\hat{r}}_{ij}:\overrightarrow{\overrightarrow{\mathfrak{q}}}_{j,2}\right)+\frac{1}{2}\mathfrak{c}_{j,0}\left(\hat{\hat{r}}_{ij}:\overrightarrow{\overrightarrow{\mathfrak{q}}}_{i,2}\right)\right]\label{Eq:CoefficientTerm2+_}
\end{equation}

\begin{equation}
c_{ij,2^{-}}\left(\overrightarrow{\mathfrak{p}}_{ij,1},\mathfrak{q}_{ij,0}\right)=\frac{m}{r_{ij}^{2}}\left[-\left(\overrightarrow{\mathfrak{p}}_{i,1}\cdot\overrightarrow{\mathfrak{p}}_{j,1}\right)+\frac{1}{2}\mathfrak{c}_{i,0}\mathfrak{q}_{j,0}+\frac{1}{2}\mathfrak{c}_{j,0}\mathfrak{q}_{i,0}\right]\label{Eq:CoefficientTerm2-_}
\end{equation}
Of course, keep in mind the linear combination of Eq.\ \ref{Eq:CoefficientTerm2},
which we cast as follows here:

\begin{equation}
c_{ij,2}\left(\overrightarrow{\mathfrak{p}}_{ij,1},\left(\overrightarrow{\overrightarrow{\mathfrak{q}}}_{ij,2},\mathfrak{q}_{ij,0}\right)\right)=\left(m+2\right)c_{ij,2^{+}}-c_{ij,2^{-}}\label{Eq:CoefficientTerm2_}
\end{equation}
Besides, note that the ``hats'' above $r$ denote a distance tensor
of a unity magnitude; their amount is equivalent with the order of
the tensor. While usually omitting yet implying the functionality
of $c_{ij,\aleph}$ in the remainder of our mathematics, we emphasize
here that the coefficients above are actually dependent on the set
of dipoles $\overrightarrow{\mathfrak{p}}_{ij,1}$, together with
the set of quadrupoles $\overrightarrow{\overrightarrow{\mathfrak{q}}}_{ij,2}$
and $\mathfrak{q}_{ij,0}$; of course, these are also functions of
$r_{ij}$, as well as of the monopoles. Besides, note that for $m=1$,
the correspondence of these expressions with the familiar Coulombic
case becomes apparent, considering our definitions of the quadruples
in Eqs.\ \ref{Eq:Quadrupole+} and \ref{Eq:Quadrupole-}. 

Importantly in our general formalism, these coefficients can be employed
in Eq.\ \ref{Eq:EnergyD} for evaluating the interaction between
the molecular pair of Fig.\ \ref{Fig:MPIL}: These can be used as
correction terms for Eq.\ \ref{Eq:Energy0_}, and for cases in which
the zero-order term vanishes, $c_{ij,1}$ or $c_{ij,2}$ may even
serve as the MPIL. Let us now consider the computational implementation
of Eqs.\ \ref{Eq:CoefficientTerm1_} and \ref{Eq:CoefficientTerm2_},
especially in relation with our corresponding discussion of Eqs.\ \ref{Eq:CoefficientTerm1}
and \ref{Eq:CoefficientTerm2}, respectively (i.e., the discussion
before we assumed the product rule of Eq.\ \ref{Eq:Coefficient}).
Above all, Eqs.\ \ref{Eq:CoefficientTerm1_} and \ref{Eq:CoefficientTerm2_}
are dependent on parameters of gravitational centers $i$ and $j$
(i.e., $\overrightarrow{\mathfrak{p}}_{ij,1}$, as well as $\overrightarrow{\overrightarrow{\mathfrak{q}}}_{ij,2}$
and $\mathfrak{q}_{ij,0}$), while Eqs.\ \ref{Eq:CoefficientTerm1}
and \ref{Eq:CoefficientTerm2} are dependent on parameters of atomistic
coordinates $\mu_{i}$ and $\nu_{j}$ (i.e., $\xi_{\mu_{i}\nu_{j}}$,
as well as $\cos\theta_{\mu_{i}\nu_{j}}$). Of course, the latter
requires a summation over $n_{i}n_{j}$ sites, while the former does
not; besides, $\overrightarrow{\mathfrak{p}}_{ij,1}$, together with
$\overrightarrow{\overrightarrow{\mathfrak{q}}}_{ij,2}$ and $\mathfrak{q}_{ij,0}$,
can be evaluated analytically at no computational cost irrespective
of $\overrightarrow{r}_{ij}$, while $\xi_{\mu_{i}\nu_{j}}$ together
$\cos\theta_{\mu_{i}\nu_{j}}$ must be calculated at each step of
a molecular simulation with consideration of $\overrightarrow{r}_{ij}$.
Still, just as with Eqs.\ \ref{Eq:CoefficientTerm1} and \ref{Eq:CoefficientTerm2},
the energy corresponding with Eqs.\ \ref{Eq:CoefficientTerm1_} and
\ref{Eq:CoefficientTerm2_} involves vectors, which is unlike the
very convenient energy function of the monopole-monopole term of Eq.\ \ref{Eq:Energy0_},
which solely depends on scalars. In turn on a practical level, we
do not recommend going beyond the zero-order term in the multipole
expansion unless it is necessary (e.g., water). 

\subsection{\textit{Defining Relative Resolution }\textit{\normalsize{}\label{SecB:Theoretical5}}}

\begin{figure}[H]
\begin{centering}
\includegraphics{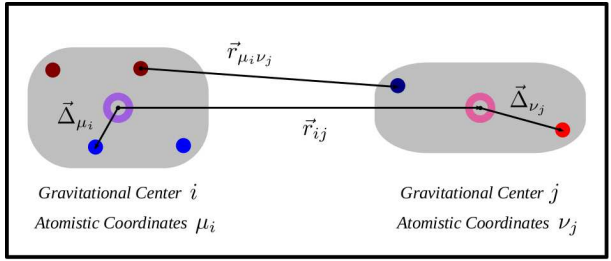}
\par\end{centering}
\caption{\textit{\label{Fig:MPIL} }A pair of molecules in vacuum. The gravitational
centers are represented by hollow rings, and the atomistic coordinates
are represented by replete disks. The left side is for molecule $i$
with its $\mu_{i}$ sites, and the right side is for molecule $j$
with its $\nu_{j}$ sites. The various colors of these mean that they
can all be inherently different in terms of their interaction parameters.
The gray shading does not have much of a physical meaning: It just
helps us delineate the approximate boundary of a molecule. The various
arrows are distance vectors. $\protect\overrightarrow{r}_{ij}$ is
the distance between centers $i$ and $j$, and $\vec{r}_{\mu_{i}\nu_{j}}$
is the distance between coordinates $\mu_{i}$ and $\nu_{j}$. Within
each molecule, $\protect\overrightarrow{\Delta}_{\mu_{i}}$ or $\protect\overrightarrow{\Delta}_{\nu_{j}}$
is a vector between an atomistic coordinate, $\mu_{i}$ or $\nu_{j}$,
and a gravitational center, $i$ or $j$, respectively. }
\end{figure}

\hspace{2em}Up until now, we have been essentially looking only at
the infinite limit associated with Fig.\ \ref{Fig:MPIL}. How do
we deal with an arbitrary distance between the molecular pair? RelRes
can resolve this issue, yet we must begin by clearly defining the
FG and CG potentials, which mostly amounts to introducing the appropriate
labels on the potentials we have been working with above. Considering
Fig.\ \ref{Fig:MPIL}, the FG potential $u_{\mu_{i}\nu_{j}}^{FG}$,
only a function of $r_{\mu_{i}\nu_{j}}$, is the fundamental interaction
between atomistic coordinates $\mu_{i}$ and $\nu_{j}$, and the CG
potential $u_{ij}^{CG}$, notably a function of $r_{ij}$, is the
apparent interaction between gravitational centers $i$ and $j$;
the latter is an approximation of the appropriate summation of the
former: 

\begin{equation}
u_{ij}^{CG}\left(\overrightarrow{r}_{ij};\left\{ \overrightarrow{\Delta}_{\mu_{i}}\right\} ,\left\{ \overrightarrow{\Delta}_{\nu_{j}}\right\} \right)\approx\sum_{\mu_{i}\nu_{j}}{\displaystyle u_{\mu_{i}\nu_{j}}^{FG}\left(r_{\mu_{i}\nu_{j}}\right)}\label{Eq:EnergyBase0}
\end{equation}
Compare this expression with Eq.\ \ref{Eq:EnergyBase}; importantly,
the right sides but not the left sides are identical between the two
expressions (i.e., $u_{\mu_{i}\nu_{j}}^{FG}=u_{\mu_{i}\nu_{j}}$ but
$u_{ij}^{CG}\neq u_{ij}$), which is in fact the reason for the approximation
in Eq.\ \ref{Eq:EnergyBase0}. With an inverse assumption analogous
with Eq.\ \ref{Eq:InversePower}, we can clearly define the approximation
here via the multipole expansion we have performed above (i.e., from
Eq.\ \ref{Eq:EnergyA} to Eq.\ \ref{Eq:EnergyC}). The following
then ensues:

\begin{equation}
{\displaystyle u_{\mu_{i}\nu_{j}}^{FG}\left(r_{\mu_{i}\nu_{j}}\right)}=\frac{c_{\mu_{i}\nu_{j}}}{r_{\mu_{i}\nu_{j}}^{m}}\label{Eq:FG}
\end{equation}

\begin{equation}
u_{ij}^{CG}\left(r_{ij};\xi_{\mu_{i}\nu_{j}},\cos\theta_{\mu_{i}\nu_{j}}\right)=\frac{c_{ij,\aleph^{*}}\left(\xi_{\mu_{i}\nu_{j}},\cos\theta_{\mu_{i}\nu_{j}}\right)}{r_{ij}^{m}}\label{Eq:CG}
\end{equation}
Remember of course the relation between the coefficients which appear
here as given by Eq.\ \ref{Eq:CoefficientTerm}, and $\aleph^{*}$
is the leading (nonvanishing) term in the Taylor series of Eq.\ \ref{Eq:TaylorTerm};
this is in fact equivalent with the MPIL. Besides, if we invoke the
product assumption of Eq.\ \ref{Eq:Coefficient}, we can cast these
potentials in terms of the ensuing multipoles (e.g., Eq.\ \ref{Eq:CoefficientTerm0_}).

\begin{figure}[H]
\begin{centering}
\includegraphics{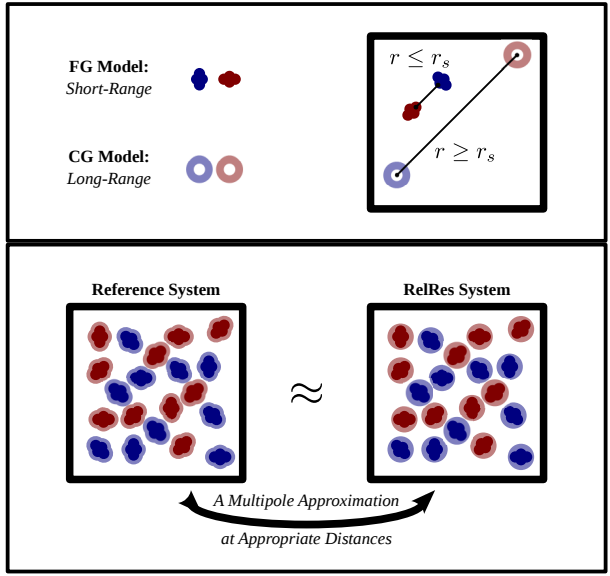}
\par\end{centering}
\caption{\textit{\label{Fig:RelRes} }A schematic representation of our multiscale
approach for nonpolar systems. The red and blue colors mean that we
formally have two different molecules here. The top panel characterizes
the geometrical FG and isotropic CG models on its left, and on its
right, it illustrates that the former applies between atomistic coordinates,
if their relative separation is small, and the latter applies between
gravitational centers, if their relative separation is large. The
bottom panel basically shows two molecular simulations, for the reference
system on the left and for the RelRes system on the right. The arrow
represents the MPIL parameterization, which makes the two systems
approximately equal to each other. }
\end{figure}

Let us now cast the interaction between the molecular pair of Fig.\ \ref{Fig:MPIL}
as a single function by using Eqs.\ \ref{Eq:FG} and \ref{Eq:CG}.
For this purpose, we must again remember the MPIL that we thoroughly
discussed above. If the molecules of Fig.\ \ref{Fig:MPIL} are near
to each other, the approximation of MPIL is unreasonable, and the
FG potential of Eq.\ \ref{Eq:FG} is the relevant one. If the molecules
of Fig.\ \ref{Fig:MPIL} are far from each other, the approximation
of MPIL is legitimate, and the CG potential of Eq.\ \ref{Eq:CG}
is the relevant one. We illustrate these ideas for nonpolar molecules
in the top panel of Fig.\ \ref{Fig:RelRes}. This is in fact the
main idea of RelRes, and we define its potential as follows:

\begin{equation}
\tilde{u}\left(r_{\mu_{i}\nu_{j}};r_{ij}\right)=\sum_{\mu_{i}\nu_{j}}{\displaystyle \tilde{u}_{\mu_{i}\nu_{j}}^{FG}\left(r_{\mu_{i}\nu_{j}}\right)+\tilde{u}_{ij}^{CG}\left(r_{ij}\right)}\label{Eq:RelRes_}
\end{equation}
In essence, RelRes is a linear combination of $\tilde{u}_{\mu_{i}\nu_{j}}^{FG}$
and $\tilde{u}_{ij}^{CG}$, which are slight modifications of the
FG and CG potentials of Eqs.\ \ref{Eq:FG} and \ref{Eq:CG}, respectively;
we thoroughly discuss these functions below. Besides, note that we
emphasize in Eq.\ \ref{Eq:RelRes_} that the RelRes potential is
a function of various pairwise distances (i.e., atomistic coordinates,
$r_{\mu_{i}\nu_{j}}$, and gravitational centers, $r_{ij}$); in such
a way, RelRes maintains all degrees of freedom. Note also that the
dependence on $\xi_{\mu_{i}\nu_{j}}$ and $\cos\theta_{\mu_{i}\nu_{j}}$
has been omitted here for clarity, especially since these variables
do not have any effect on nonpolar molecules. 

The modifications of the FG and CG potentials must ensure that for
small relative separations, $\tilde{u}_{\mu_{i}\nu_{j}}^{FG}=u_{\mu_{i}\nu_{j}}^{FG}$
while the derivative of $\tilde{u}_{ij}^{CG}$ is zero, and for large
relative separations, $\tilde{u}_{\mu_{i}\nu_{j}}^{CG}=u_{\mu_{i}\nu_{j}}^{CG}$
while the derivative of $\tilde{u}_{ij}^{FG}$ is zero. Their exact
functionalities are not unique, yet we choose the following piecewise
functions for our purposes: 

\begin{equation}
\tilde{u}_{\mu_{i}\nu_{j}}^{FG}\left(r\right)=\begin{cases}
\begin{array}{c}
u_{\mu_{i}\nu_{j}}^{FG}\left(r\right)-u_{\mu_{i}\nu_{j}}^{FG}\left(r_{s}\right)\\
0
\end{array} & \begin{array}{c}
\mathrm{if}\quad r\leq r_{s}\\
\mathrm{if}\quad r\geq r_{s}
\end{array}\end{cases}\label{Eq:FG_}
\end{equation}

\begin{equation}
\tilde{u}_{ij}^{CG}\left(r\right)=\begin{cases}
\begin{array}{c}
u_{ij}^{CG}\left(r_{s}\right)\\
u_{ij}^{CG}\left(r\right)
\end{array} & \begin{array}{c}
\mathrm{if}\quad r\leq r_{s}\\
\mathrm{if}\quad r\geq r_{s}
\end{array}\end{cases}\label{Eq:CG_}
\end{equation}
The switching distance $r_{s}$ is a constant that represents the
relative separation at which the MPIL supposedly switches from being
unreasonable to being legitimate. Note that the vertical shifts $u_{\mu_{i}\nu_{j}}^{FG}\left(r_{s}\right)$
and $u_{ij}^{CG}\left(r_{s}\right)$ merely ensure that the corresponding
forces are continuous throughout the domains; through the linear combination
of RelRes, these vertical shifts approximately cancel each other out,
with the extent of the cancellation depending on how many nonzero
terms the summation of Eq.\ \ref{Eq:RelRes_} includes. 

Eq.\ \ref{Eq:RelRes_} is actually the pairwise version of RelRes,
which only applies for a molecular pair in vacuum. Extending RelRes
for a multiscale simulation with many molecules, we define the complete
version of RelRes through a summation over all different pairs of
molecules in the system: 

\[
\tilde{U}=\frac{1}{2}\sum_{i\neq j}\tilde{u}\left(r_{\mu_{i}\nu_{j}};r_{ij}\right)
\]

\begin{equation}
\tilde{U}=\frac{1}{2}\sum_{i\neq j}\left[\sum_{\mu_{i}\nu_{j}}{\displaystyle \tilde{u}_{\mu_{i}\nu_{j}}^{FG}\left(r_{\mu_{i}\nu_{j}}\right)+\tilde{u}_{ij}^{CG}\left(r_{ij}\right)}\right]\label{Eq:RelRes}
\end{equation}
We omit the functionality of this complete version of RelRes throughout
our work for clarity. This Hamiltonian introduces a notable conceptual
complexity, as compared with Eq.\ \ref{Eq:RelRes_} of the pairwise
case. For a molecular pair in vacuum, the entire system is basically
either a pure FG or CG scenario: If the molecules are near to each
other, they are both described by the FG model, and if the molecules
are far from each other, they are both described by the CG model (of
course, there are subtleties for moderate relative separations). Nevertheless,
in the case of a molecular simulation of a realistic liquid, the situation
is completely different, with the system being neither purely FG nor
CG, but instead having a hybrid nature: Each molecule simultaneously
embodies both models, and a given molecule interacts with its near
neighbors via the FG potential and with its far neighbors via the
CG potential. The bottom panel of Fig.\ \ref{Fig:RelRes} illustrates
a RelRes system for a nonpolar scenario, which emphasizes that all
molecules embody both molecular resolutions. 

Let us also discuss RelRes in terms of its $r_{s}$ parameter, labeling
such a dependence as $\tilde{U}_{r_{s}}$. The two limits of $r_{s}$
are of particular importance for emphasis: $\tilde{U}_{\infty}$ is
identical with the corresponding Hamiltonian of the pure FG system,
and $\tilde{U}_{0}$ is identical with the corresponding Hamiltonian
of the pure CG system. The former is the basis for reference mixtures.
Such a system has $r_{s}\rightarrow\infty$ (i.e., a system with just
FG and no CG interactions), and it is depicted in the bottom panel
of Fig.\ \ref{Fig:RelRes}. An inherent goal for RelRes with an arbitrary
$r_{s}$ is of approximately capturing all behavior of its corresponding
reference mixture (e.g., $\tilde{U}_{r_{s}}\approx\tilde{U}_{\infty}$).
For completeness, let us also define the entire system Hamiltonian: 

\begin{equation}
\tilde{E}=K+U_{0}+\tilde{U}\label{Eq:Hamiltonian}
\end{equation}
$K$ is the kinetic energy, while $U_{0}$ accounts for all intramolecular
interactions. Note that these do not have a tilde above them. This
is because RelRes maintains all degrees of freedom, and in turn, its
$K$, as well as its $U_{0}$, is strictly unaltered with the corresponding
functions of the reference system (e.g., $K_{r_{s}}=K_{\infty}$). 

\pagebreak{}
\begin{singlespace}

\section{\textit{Computational Validation \label{SecA:Computational}}}
\end{singlespace}

We now turn our attention for testing the efficacy of our multiscale
approach in describing reference systems . We do this via molecular
simulations, complementing in turn the initial examination of our
previous publication \citep{ChaimovichKremer_JCP2015}. Our original
work was a proof of concept that RelRes is very successful in describing
(nonpolar) multi-component and multi-phase systems across state space;
by invoking a tuning parameter in our Hamiltonian, we also showed
that MPIL is in fact the ideal choice for the potential beyond nearest
neighbors. Consequently for simplicity in this current work, we only
examine uniform liquids, taking MPIL for granted in all of our systems.
Here, we instead focus on systematically varying the distance at which
we switch between the FG and CG potentials. Besides, we also complement
our initial work by showing that RelRes with MPIL works well not only
for static behavior but also for dynamic behavior. 

\vspace{4ex}

\subsection{\textit{Constructing the Molecular Simulations \label{SecB:Computational1} }}

\hspace{2em}Let us begin by thoroughly discussing the general implementation
of our molecular simulations. It is almost identical with the implementation
we had in our original publication \citep{ChaimovichKremer_JCP2015};
we emphasize important aspects which are different. Above all, we
again perform our molecular simulations with the GROMACS package,
specifically its 4.6 version \citep{HessLindahl_JCTC2008}. If we
do not specify here certain features of the molecular simulations,
that basically means that the default values are used. 

In the current publication, all of our systems contain a total of
2000 (identical) molecules: All of them are dumbbell-like molecules
(e.g., ethenes), with two FG sites and one CG site, meaning that $n=2$.
All FG atomistic coordinates have mass $m$, and all CG gravitational
centers have mass zero; the latter is constructed via equal weights
of the former. 

In all cases, the LJ potential is the sole component for the interactions
between molecules; this in turn sets the length and energy parameters,
$\sigma$ and $\epsilon$, respectively, for our work. For a practical
implementation of RelRes so that there are no singularities in Newtonian
trajectories, we modify the step functionality of Eqs.\ \ref{Eq:FG_}
and \ref{Eq:CG_}: In each of these in the interval $\left[r_{s}-\delta r_{s},r_{s}+\delta r_{s}\right]$,
we introduce the following four-term polynomial in terms of the inverse
distance: 

\begin{equation}
f\left(r\right)=\frac{\gamma_{6}}{r^{6}}+\frac{\gamma_{12}}{r^{12}}+\frac{\gamma_{18}}{r^{18}}+\frac{\gamma_{24}}{r^{24}}\label{Eq:Sigmoid}
\end{equation}
The four coefficients $\gamma$ are determined by ensuring that the
potential, as well as its derivative, is continuous at its two boundaries,
$r_{s}\pm\delta r_{s}$. We vary $r_{s}$ between $1.1\sigma$ and
$1.9\sigma$ at intervals of $0.2\sigma$, and we set $\delta r_{s}=0.0625\sigma$.
On an analogous note, rather than sharply truncating all potentials
at $2.5\sigma$ with a step function, we invoke the sigmoid function
above between $2.25\sigma$ and $2.75\sigma$. Note that all interactions
in our reference system, denoted by $r_{s}\rightarrow\infty$, practically
vanish beyond this distance. 

The intramolecular energetics are mostly governed by elastic bonds,
which is unlike our previous publication that had all bonds being
rigid. Here, the distance of each bond is $0.5\sigma$, while its
spring is $2000\epsilon/\sigma^{2}$. Comparing with a conventional
bond, our distance is shorter by a significant fraction, while our
spring is about an order of magnitude weaker. This is purposefully
done as a stringent test: If RelRes can work on our large flexible
molecules, it can also work on small stiff molecules. 

In all cases, the protocol that we use involves a sequence of molecular
simulations at a temperature of $1.0\epsilon/k$, with $k$ being
Boltzmann\textquoteright s constant. The initial molecular simulation
is of the reference liquid, and it is distinct in that it is coupled
to a barostat, whose pressure is $1.0\epsilon/\sigma^{3}$; the purpose
of this molecular simulation is to fix the system size for the entire
sequence, while all of its results are disregarded. The rest of the
molecular simulations are subsequently in the canonical ensemble:
One of them is of the reference liquid, while the rest are RelRes
systems of different switching distances. Besides, all molecular simulations
evolve via Newtonian equations of motion. Each molecular simulation
starts with an equilibration phase of $5,000$ steps of size $0.002\tau$
and ends with a production phase of $1,000,000$ steps of size $0.001\tau$;
note that $\tau=\sqrt{m\sigma^{2}/\epsilon}$ is our unit of time.
Realize that all molecular simulations that we perform in this work
are of single-component and single-phase systems. 

\subsection{\textit{Computing Static and Dynamic Features \label{SecB:Computational2} }}

Above all, we thoroughly examine in our work several structural correlations,
usually in terms of relative separation $r_{ij}$; we frequently omit
the $ij$ indices throughout most of our work. For this purpose, we
record the positions of all molecules every $100$ steps of the molecular
simulation. The domain of our $r$ goes completely from the origin
to the edge of the system box; we discretize it by $1000$ bins. 

Foremost, we look at radial distributions, $g$, between the midpoints
of bonds. For comparison between radial distributions, it is convenient
to invoke a functional for them. We particularly choose the Jensen-Shannon
entropy: 

\begin{equation}
S_{JS}=S\left[\frac{1}{2}\left(g_{r_{s}}+g_{\infty}\right)\right]-\frac{1}{2}\left(S\left[g_{r_{s}}\right]+S\left[g_{\infty}\right]\right)\label{Eq:Entropy_JS}
\end{equation}
with the conventional entropy taking on this definition: 

\begin{equation}
S\left[g\right]=-\intop\kappa gr^{2}\ln\left(\kappa gr^{2}\right)dr\label{Eq:Entropy}
\end{equation}
$\kappa$ is just the normalization constant (i.e., $\intop\kappa gr^{2}=1$),
which is inversely proportional with the system volume. In Eq.\ \ref{Eq:Entropy_JS},
the term on the left (i.e., $S\left[\frac{1}{2}\left(g_{\infty}+g_{r_{s}}\right)\right]$)
is one entropy, altogether for an average of two distributions, and
the term on the right (i.e., $\frac{1}{2}\left(S\left[g_{\infty}\right]+S\left[g_{r_{s}}\right]\right)$)
is an average of two entropies, each for one distribution. As such,
the Jensen-Shannon entropy is a functional of two radial distributions,
which just measures the disparity between them: Obviously if $g_{r_{s}}\approx g_{\infty}$,
$S_{JS}\approx0$, and $S_{JS}$ increases as the discrepancy between
$g_{r_{s}}$ and $g_{\infty}$ intensifies. 

Note that while in the previous work, we presented $g$ between atomistic
coordinates, in the current work, we basically focus on $g$ between
gravitational centers. In turn, it is important to complement these
radial distributions with orientational correlations. By letting a
bond be a vector, we define $\vec{s}_{i}$ and $\vec{s}_{j}$ as the
respective directions of molecules $i$ and $j$. We particularly
focus on computing the moments of their corresponding dot product,
$\vec{s}_{i}\cdot\vec{s}_{j}$, as a function of the relative separation
between the midpoints of the bonds. Most of our case studies have
symmetry in their molecules, and thus for $\vec{s}_{i}\cdot\vec{s}_{j}$,
the average vanishes but the variance persists. We consequently present
in our work only the latter, $\left\langle \left(\vec{s}_{i}\cdot\vec{s}_{j}\right)^{2}\right\rangle $;
note that $\left\langle \left(\vec{s}_{i}\cdot\vec{s}_{j}\right)^{2}\right\rangle =\frac{1}{3}$
for decorrelated cases. Moreover, the bond vectors can be cast as
a linear decomposition in terms of two components $\vec{s}_{i}=\vec{s}_{i}^{\Vert}+\vec{s}_{i}^{\bot}$
and $\vec{s}_{j}=\vec{s}_{j}^{\Vert}+\vec{s}_{j}^{\bot}$: In terms
of the relative separation between molecules $i$ and $j$, $\left\{ \vec{s}_{i}^{\Vert},\vec{s}_{j}^{\Vert}\right\} $
are the components parallel to $\vec{r}_{ij}$, and $\left\{ \vec{s}_{i}^{\bot},\vec{s}_{j}^{\bot}\right\} $
are the components perpendicular to $\vec{r}_{ij}$. In turn, we have
the following:

\[
\left\langle \left(\vec{s}_{i}\cdot\vec{s}_{j}\right)^{2}\right\rangle =\left\langle \left(s_{i}^{\Vert}s_{j}^{\Vert}\right)^{2}\right\rangle +2\left\langle s_{i}^{\Vert}s_{j}^{\Vert}s_{i}^{\bot}s_{j}^{\bot}\right\rangle +\left\langle \left(s_{i}^{\bot}s_{j}^{\bot}\right)^{2}\right\rangle 
\]

\begin{equation}
\left\langle \left(\vec{s}_{i}\cdot\vec{s}_{j}\right)^{2}\right\rangle \approx\left\langle \left(s_{i}^{\Vert}s_{j}^{\Vert}\right)^{2}\right\rangle +\left\langle \left(s_{i}^{\bot}s_{j}^{\bot}\right)^{2}\right\rangle \label{Eq:Directions}
\end{equation}
The approximation is introduced here because in most of our case studies,
we observe that the cross term in the above expression is rather negligible,
and thus, we only present the two main components, $\left\langle \left(s_{i}^{\Vert}s_{j}^{\Vert}\right)^{2}\right\rangle $
and $\left\langle \left(s_{i}^{\bot}s_{j}^{\bot}\right)^{2}\right\rangle $,
of this orientational function; realize that $\left\langle \left(s_{i}^{\Vert}s_{j}^{\Vert}\right)^{2}\right\rangle =\frac{1}{9}$
and $\left\langle \left(s_{i}^{\bot}s_{j}^{\bot}\right)^{2}\right\rangle =\frac{2}{9}$
for an ideal gas. 

Besides these static functions, we also examine dynamic functions,
something that we did not at all do in our original publication. Specifically,
we look at the squared displacement $\left\langle \left(\vec{r}_{t}-\vec{r}_{0}\right)^{2}\right\rangle $,
as well as its derivative, $\partial_{t}\left\langle \left(\vec{r}_{t}-\vec{r}_{0}\right)^{2}\right\rangle $.
$\vec{r}_{t}$ is the position of molecule $i$ or $j$ at a given
time $t$, and the index $0$ is of course for zero time. Besides,
we also examine an orientational correlation $\left\langle \vec{s}_{t}\cdot\vec{s}_{0}\right\rangle $.
$\vec{s}_{t}$ is analogous with the dumbbell directions introduced
earlier; just as with the squared displacement, we have here the time
index, while the molecule indices, $i$ or $j$, are omitted for clarity.
Importantly in calculating these transport functions, we employ an
algorithm of multiple time origins, in fact using each recorded step
for such purposes. 

We furthermore examine thermal properties in our work. We foremost
look at the average $\left\langle \widetilde{U}\right\rangle $ and
variance $\left\langle \delta\widetilde{U}^{2}\right\rangle $ of
the defining equation of RelRes. For convenience, we normalize both
by the corresponding values for the reference liquid: $\left\langle \widetilde{U}\right\rangle _{r_{s}}^{*}=\left\langle \widetilde{U}\right\rangle _{r_{s}}/\left\langle \widetilde{U}\right\rangle _{\infty}$
and $\left\langle \delta\widetilde{U}^{2}\right\rangle _{r_{s}}^{*}=\left\langle \delta\widetilde{U}^{2}\right\rangle _{r_{s}}/\left\langle \delta\widetilde{U}^{2}\right\rangle _{\infty}$;
as we frequently do, we omit the index of $r_{s}$ for these thermal
properties. Besides, we also look at a transport function of the total
energy, $\left\langle \delta\widetilde{E}_{t}\delta\widetilde{E}_{0}\right\rangle /\left\langle \delta\widetilde{E}^{2}\right\rangle $,
computing it again using multiple time origins. All of our thermal
properties are based on probing our molecular simulations every $10$
steps. 

\subsection{\textit{Examining Elementary Dumbbells }\textit{\normalsize{}\label{SecB:Computational3}}}

\begin{figure}[H]
\begin{centering}
\includegraphics{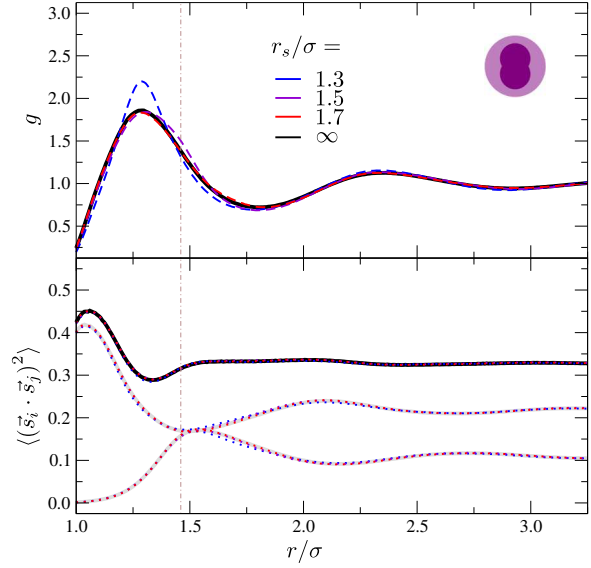}
\par\end{centering}
\caption{\textit{\label{Fig:BasisStc} }Static correlations for the elementary
dumbbell scenario. These molecules have uniform LJ parameters for
their two sites, $\sigma$ and $\epsilon$, and the bond length between
them is $0.5\sigma$. Here, all the black curves, as well as the gray
ones, correspond with the reference system. The colored curves, correspond
with the RelRes strategy, with each color representing a different
switching distance $r_{s}$. Note particularly the violet curve, which
is of the ideal switching distance in capturing the behavior of the
reference liquid; the blue or red curve has a shorter or longer switching
distance, respectively. Everything is plotted as a function of the
distance between the midpoints of the dumbbells, $r$. The top panel
plots the corresponding radial distribution $g$, with the solid curve
being for the reference system, while the dashed curves being for
the RelRes scenarios. The bottom panel, for the reference liquid,
plots the orientational correlation $\left\langle \left(\vec{s}_{i}\cdot\vec{s}_{j}\right)^{2}\right\rangle $
as the black curve, together with its two main components $\left\langle \left(s_{i}^{\Vert}s_{j}^{\Vert}\right)^{2}\right\rangle $
and $\left\langle \left(s_{i}^{\bot}s_{j}^{\bot}\right)^{2}\right\rangle $
as the lower and higher gray curves, respectively; the functionality
of the dotted curves corresponds with the functionality of their neighboring
solid curves; note that all of these curves are almost identical.
In both panels, the vertical brown line represents the inflection
of $g$, which is located here at $1.46$. }
\end{figure}

\hspace{2em}We begin by examining a system of elementary dumbbells.
These can be considered as ethenes, and one such molecule is depicted
in Fig.\ \ref{Fig:BasisStc}. Such a dumbbell has two identical FG
sites, with $\sigma$ and $\epsilon$ for its LJ parameters. By the
MPIL, this means that the respective CG site has $\sigma$ and $4\epsilon$
for its LJ parameters since $n=2$. The inherent bond of these elementary
dumbbells is set at at a distance of $0.5\sigma$. The density for
this system is $1.00\sigma^{3}/m$. 

The radial distribution between the dumbbells is given in the top
panel of Fig.\ \ref{Fig:BasisStc}. This structural correlation for
the reference liquid is given here as the solid black curve; the remaining
dashed curves are for the RelRes systems, with each color representing
a different switching distance. All of these switching distances give
a sufficient description of the radial distribution of the reference
system, and it is clear that as $r_{s}$ increases, the capability
of RelRes improves, in an apparently asymptotic manner: While for
$r_{s}=1.3\sigma$ (i.e., the blue curve), the replication is fairly
decent, for $r_{s}=1.7\sigma$ (i.e., the red curve), the replication
is essentially perfect; in fact, $r_{s}=1.5\sigma$ (i.e., the violet
curve) already captures the radial distribution of the reference system
as well as one may desire. The systematic observation here is reminiscent
with the one we had in the original publication \citep{ChaimovichKremer_JCP2015}:
RelRes captured structural correlations with $r_{s}\approx1.1\sigma$
equitably yet with $r_{s}\approx1.6\sigma$ splendidly. 

\begin{figure}[H]
\begin{centering}
\includegraphics{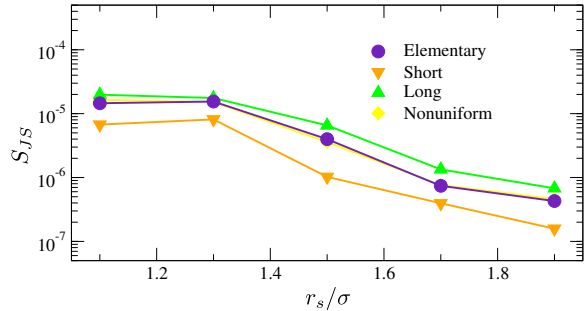}
\par\end{centering}
\caption{\textit{\label{Fig:EntropyDmbl} }The Jensen-Shannon entropy, $S_{JS}$,
for the radial distribution, plotted as a function of the switching
distance $r_{s}$ in RelRes. Each curve is for a different system.
The indigo color is for the molecules which have identical LJ parameters
between their two sites, $\epsilon$ and $\sigma$, and the corresponding
bond length is $0.5\sigma$. The orange and green curves refer to
the systems with bond lengths of $0.3\sigma$ and $0.7\sigma$, while
their LJ parameters are also uniform. The almost hidden yellow curve
corresponds with $\left\{ 1.61\epsilon,0.62\epsilon\right\} $ and
$\left\{ 0.85\sigma,1.18\sigma\right\} $ for its LJ parameters, while
its bond is identical with that of the base case. Importantly, note
that the entropy is in logarithmic scale. Besides, the lines here
just serve as guides. }
\end{figure}

For the purpose of better clarifying this asymptotic behavior, we
invoke the Jensen-Shannon entropy, as defined by Eq.\ \ref{Eq:Entropy_JS}.
For each of the relevant curves in Fig.\ \ref{Fig:BasisStc}, we
calculate $S_{JS}$, and we plot it (i.e., an indigo circle) in terms
of the switching distance in Fig.\ \ref{Fig:EntropyDmbl}. Clearly,
this entropy is monotonic throughout most of the domain of $r_{s}$,
and considering the logarithmic scale here, $S_{JS}$ is roughly an
exponential decay of the distance; this is especially true around
$r_{s}=1.5\sigma$, which we mentioned in the context of Fig.\ \ref{Fig:BasisStc}
as the critical switching distance for adequately capturing the radial
distribution of the reference system. Thus, we confirm here our earlier
observation: RelRes improves asymptotically as $r_{s}$ approaches
infinity. Importantly, it appears that the asymptote is practically
reached at $r_{s}\approx1.5\sigma$: At this distance, the discrepancy
between the radial distributions almost vanishes, and in turn, $S_{JS}$
becomes rather negligible.

So why is $1.5\sigma$ the critical distance for switching between
the FG and CG potentials in RelRes? For this purpose, we now look
at orientational correlations in the bottom panel of Fig.\ \ref{Fig:BasisStc}.
Let us begin by discussing the black solid curve, which is $\left\langle \left(\vec{s}_{i}\cdot\vec{s}_{j}\right)^{2}\right\rangle $
of the reference liquid. Notice that while it clearly has a maximum
and a minimum early on, it quickly flattens out after the ensuing
inflection at $1.42\sigma$. This consequently explains why $1.5\sigma$
is an excellent choice for $r_{s}$ in RelRes: At this distance, the
dumbbell directions apparently become decorrelated, and the influence
of these orientational degrees of freedom becomes negligible on the
various structural correlations of the system (e.g., radial distributions).
Interestingly, the flattening of $\left\langle \left(\vec{s}_{i}\cdot\vec{s}_{j}\right)^{2}\right\rangle $
in the bottom panel happens just around the middle between the respective
maximum and minimum in $g$ of the corresponding top panel (i.e.,
the relative separation of $1.54\sigma$); this is also around the
respective inflection of $g$ (i.e., the relative separation of $1.46\sigma$),
which we mark by a vertical brown line in both panels. This means
that as soon as nearest neighbors depart from each other, their directions
quickly become decorrelated. 

Let us continue examining the two main components of $\left\langle \left(\vec{s}_{i}\cdot\vec{s}_{j}\right)^{2}\right\rangle $,
as given in Eq.\ \ref{Eq:Directions}. For the reference system,
these are plotted as gray curves in the bottom panel of Fig.\ \ref{Fig:BasisStc},
the lower one is for $\left\langle \left(s_{i}^{\Vert}s_{j}^{\Vert}\right)^{2}\right\rangle $
and the higher one is for $\left\langle \left(s_{i}^{\bot}s_{j}^{\bot}\right)^{2}\right\rangle $;
importantly, realize that their summation essentially yields the black
curve. Surprisingly, the behavior of $\left\langle \left(\vec{s}_{i}\cdot\vec{s}_{j}\right)^{2}\right\rangle $
is very distinct from its components: While $\left\langle \left(\vec{s}_{i}\cdot\vec{s}_{j}\right)^{2}\right\rangle $
becomes mostly decorrelated as the molecules leave the primary coordination
shell, its components are still correlated even as the molecules enter
the secondary coordination shell. In fact, $\left\langle \left(s_{i}^{\Vert}s_{j}^{\Vert}\right)^{2}\right\rangle $
and $\left\langle \left(s_{i}^{\bot}s_{j}^{\bot}\right)^{2}\right\rangle $
appear as mirror images of each other, especially beyond $r\approx1.5\sigma$,
with the fluctuations between the maxima and minima of the two curves
occurring at the same locations. In turn, once we sum these two functions,
the extrema cancel each other, yielding the overall function which
is basically flat beyond $r\approx1.5\sigma$. The locations of the
initial maximum of $\left\langle \left(s_{i}^{\Vert}s_{j}^{\Vert}\right)^{2}\right\rangle $
and the initial minimum of $\left\langle \left(s_{i}^{\bot}s_{j}^{\bot}\right)^{2}\right\rangle $
occur at $1.49\sigma$ and $1.54\sigma$, which is again almost identical
with the ideal $r_{s}$ of $1.5\sigma$. 

Finally, let us not forget the dotted curves in the bottom panel of
Fig.\ \ref{Fig:BasisStc}. Their coloring is equivalent with that
of the corresponding top panel, with each color representing a different
$r_{s}$ in RelRes; besides, the function to which each dotted curve
pertains corresponds with the function of its neighboring solid curve.
Importantly, RelRes captures the orientational correlation with its
components perfectly well, regardless of which switching distances
we use. This is in contrast with the radial distribution, which is
noticeably influenced by $r_{s}$. 

\begin{figure}[H]
\begin{centering}
\includegraphics{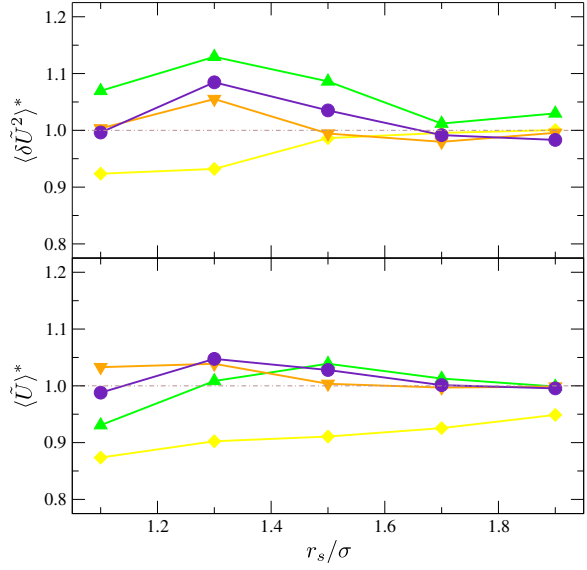}
\par\end{centering}
\caption{\textit{\label{Fig:EnergyDmbl} }Energetic functionals plotted as
a function of the switching distance $r_{s}$ in RelRes. Particularly,
the average $\left\langle \widetilde{U}\right\rangle $ and variance
$\left\langle \delta\widetilde{U}^{2}\right\rangle $ of the configurational
energy is given in the bottom and top panels respectively; the asterisk
denotes that these functionals are normalized by the respective value
of the reference system. Besides, the color coding here is identical
with that of Fig.\ \ref{Fig:EntropyDmbl}. }
\end{figure}

We now move on to examine thermal properties of these elementary dumbbells.
We specifically look at functionals of the configurational Hamiltonian
of RelRes, presenting, as a function $r_{s}$, its normalized average
$\left\langle \widetilde{U}\right\rangle $ and variance $\left\langle \delta\widetilde{U}^{2}\right\rangle $
in the bottom and top panels of Fig.\ \ref{Fig:EnergyDmbl}, respectively.
The coloring here is analogous with Fig.\ \ref{Fig:EntropyDmbl}.
Focus again on the indigo circles: Because of normalization, they
fluctuate about the horizontal brown line which is of course set at
unity. In fact, for all switching distances, the average and variance
are essentially both within a fraction of $0.05$ from their reference
values. Importantly, we notice that a dampening effect of the fluctuation,
with both the average and variance eventually reaching unity. Besides,
the fluctuating characteristic of the symbols in this figure is in
striking contrast to the observation we made for the functional of
the structural correlations in Fig.\ \ref{Fig:EntropyDmbl}, which
showed an asymptotic behavior with $r_{s}$. Bearing also in mind
Fig.\ \ref{Fig:BasisStc}, this means that if one is fine with a
decent yet rough description of radial distributions in a liquid,
one can still adequately capture thermal properties, together with
orientational correlations, while using just a modest switching distance
in RelRes that does not entirely account for all nearest neighbors.
Finally, these systematic findings are analogous with the observations
we made in our previous work for the pressure, together with its corresponding
response function. For both of our switching distances there, $r_{s}\approx1.1\sigma$
and $r_{s}\approx1.6\sigma$, RelRes was very successful once we used
it together with MPIL; \citep{ChaimovichKremer_JCP2015}. In general,
the excellent replication of thermal properties reiterates the validity
of the MPIL approximation, which is essentially based on energy conservation. 

\begin{figure}[H]
\begin{centering}
\includegraphics{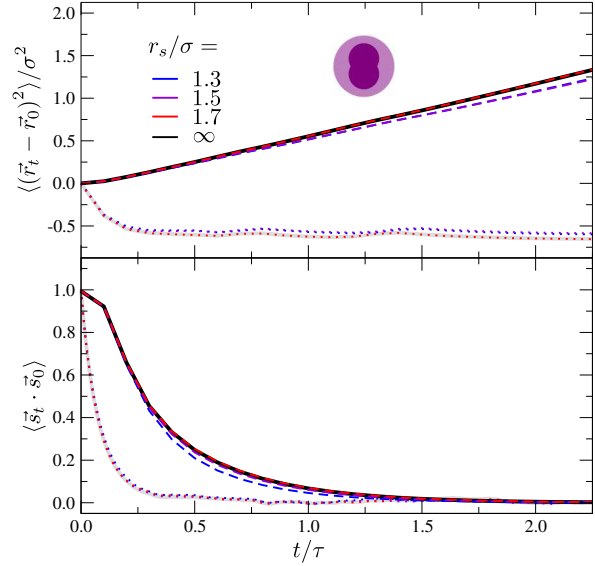}
\par\end{centering}
\caption{\textit{\label{Fig:BasisDnc}} Dynamic correlations for the elementary
dumbbell scenario. Much of the coding here is equivalent with Fig.\ \ref{Fig:BasisStc},
but all of the current functions are different from the elementary
case. Everything is plotted in terms of time $t$. The top panel plots
the squared displacement $\left\langle \left(\vec{r}_{t}-\vec{r}_{0}\right)^{2}\right\rangle $
of a given molecule as the black curve for the reference system, together
with its neighboring dashed lines for the RelRes systems; the corresponding
negative derivative of this function $-\partial_{t}\left\langle \left(\vec{r}_{t}-\vec{r}_{0}\right)^{2}\right\rangle $
is given as the gray curve, together with its dotted lines. In an
analogous manner, the bottom panel plots the orientational function
$\left\langle \vec{s}_{t}\cdot\vec{s}_{0}\right\rangle $ as the black
curve, together with its neighboring dashed lines; the gray curve,
together with its dotted lines, represents the energetic function
$\left\langle \delta\widetilde{E}_{t}\delta\widetilde{E}_{0}\right\rangle /\left\langle \delta\widetilde{E}^{2}\right\rangle $. }
\end{figure}

So far, considering our original publication as well, we have only
examined static behavior; we now turn our attention to dynamic behavior.
We consequently plot various transport functions in terms of time,
$t$, in Fig.\ \ref{Fig:BasisDnc}. The coloring here is again analogous
with Fig.\ \ref{Fig:BasisStc}: While all solid curves are for the
reference liquid, the dashed and dotted lines are for the RelRes systems
with varying $r_{s}$. In the top panel for the reference liquid,
we give $\left\langle \left(\vec{r}_{t}-\vec{r}_{0}\right)^{2}\right\rangle $
as the black line, with its negative derivative $-\partial_{t}\left\langle \left(\vec{r}_{t}-\vec{r}_{0}\right)^{2}\right\rangle $
as the gray one. In the bottom panel for the reference liquid, the
orientational correlation $\left\langle \vec{s}_{t}\cdot\vec{s}_{0}\right\rangle $
is given as the black line. Besides these structural functions, we
also present here a thermal function, particularly for the total energy,
$\left\langle \delta\widetilde{E}_{t}\delta\widetilde{E}_{0}\right\rangle /\left\langle \delta\widetilde{E}^{2}\right\rangle $,
as the gray line. Of course, the functionality of the colored lines
is the same as that of their respective neighboring curves. 

In an all cases, irrespective of the value of $r_{s}$, RelRes satisfactorily
captures the transport functions of the reference system. Of course,
there are some nuances between the various transport functions in
terms of $r_{s}$, and these dynamic characteristics are reminiscent
of our observations for the static features. For example, $r_{s}=1.7\sigma$
is required for a flawless replication of the translational $\left\langle \left(\vec{r}_{t}-\vec{r}_{0}\right)^{2}\right\rangle $,
yet for the orientational $\left\langle \vec{s}_{t}\cdot\vec{s}_{0}\right\rangle $,
only $r_{s}=1.3\sigma$ can not excellently capture the behavior of
the reference liquid. Besides, for the thermal function $\left\langle \delta\widetilde{E}_{t}\delta\widetilde{E}_{0}\right\rangle /\left\langle \delta\widetilde{E}^{2}\right\rangle $,
all switching distances yield a perfect description of this transport
function. In summary, in terms of $r_{s}$, we notice the same trends
for dynamic behavior as we do with static behavior: Thermal properties
can be superbly captured even with a surprisingly small $r_{s}$,
but structural correlations require a relatively large $r_{s}$; for
the latter, we notice that its orientational correlations are rather
feasible of capturing as compared with their translational counterparts.
Specifically for these elementary dumbbells, it appears that $r_{s}=1.5\sigma$
can capture the entire behavior of the reference liquid very well,
and thus, this is our recommended switching distance for these molecules.
This in turn reiterates one of our main findings of the original publication:
RelRes works best if molecules interact with each via a FG potential
between nearest neighbors and a CG potential between other neighbors. 

\subsection{\textit{Varying the Bond Length }\textit{\normalsize{}\label{SecB:Computational4}}}

\begin{figure}[H]
\begin{centering}
\includegraphics{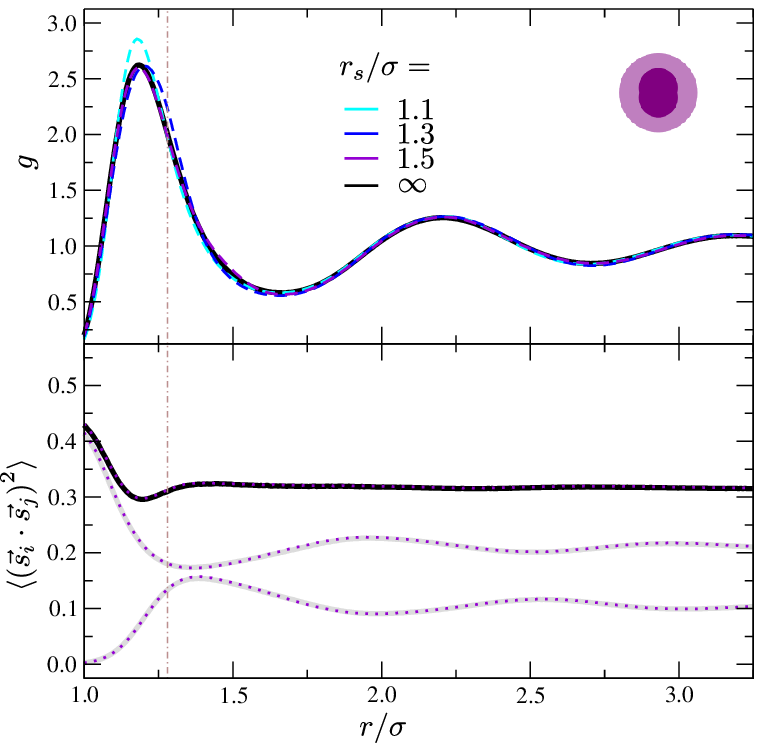}
\par\end{centering}
\caption{\textit{\label{Fig:Bnd12Stc}} Static correlations for the short dumbbells
with a bond length of $0.3\sigma$. Everything here is essentially
equivalent with Fig.\ \ref{Fig:BasisStc}. However, note the cyan
curve which is for RelRes with a switching distance of $1.1\sigma$.
The vertical line here goes through $1.28$. We plot the orientational
function only for $r_{s}=1.5\sigma$, since all other curves are basically
identical with it.}
\end{figure}

\begin{figure}[H]
\begin{centering}
\includegraphics{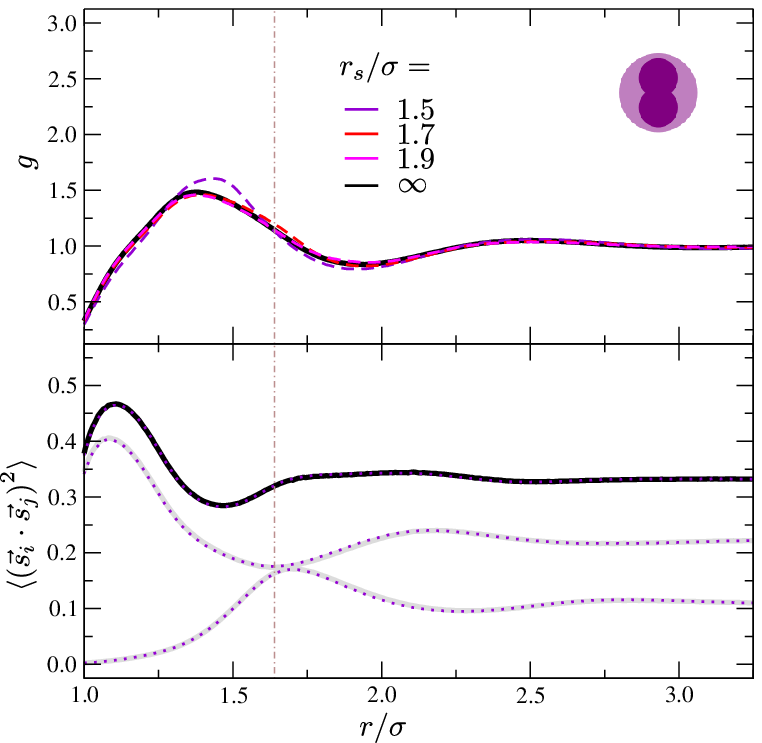}
\par\end{centering}
\caption{\textit{\label{Fig:Bnd28Stc} }Static correlations for the long dumbbells
with a bond length of $0.7\sigma$. Everything here is essentially
equivalent with Fig.\ \ref{Fig:BasisStc}. However, note the magenta
curve which is for RelRes with a switching distance of $1.9\sigma$.
The vertical line here goes through $1.64$. We plot the orientational
function only for $r_{s}=1.5\sigma$, since all other curves are basically
identical with it.}
\end{figure}

\hspace{2em}We continue by examining dumbbells in which we vary the
length of the bonds. In particular, we construct two systems, one
with short dumbbells (i.e., $\ell=0.3\sigma$) and one with long dumbbells
(i.e., $\ell=0.7\sigma$); representative sketches are given in Figs.\ \ref{Fig:Bnd12Stc}
and \ref{Fig:Bnd28Stc}, respectively. The modification of the bond
length does not alter the LJ parameters of the FG and CG sites: They
are again $\sigma$ and $\epsilon$ for the former, and $\sigma$
and $4\epsilon$ for the latter. The system density is $1.34$ for
the short molecules and $0.78$ for the long molecules, in units of
$m/\sigma^{3}$. 

The structural correlations for the short and long dumbbells are given
in Figs.\ \ref{Fig:Bnd12Stc} and \ref{Fig:Bnd28Stc}, respectively,
which are basically analogous with those of Fig.\ \ref{Fig:BasisStc}.
All orientational functions of the reference liquids are perfectly
captured by RelRes, regardless of the switching distance that we use;
we consequently give in the respective bottom panels the $r_{s}=1.5\sigma$
curve. Conversely, the radial distributions are given in the respective
top panels; notice importantly the cyan curve (i.e., $r_{s}=1.1\sigma$)
for the $\ell=0.3\sigma$ case and the magenta curve (i.e., $r_{s}=1.9\sigma$)
for the $\ell=0.7\sigma$. In essence for each case, we have shifted
our systematic $r_{s}$ examination by $0.2\sigma$. This is because
we observe that RelRes captures the radial distributions at a switching
distance that differs by roughly $0.2\sigma$ as compared with the
$\ell=0.5\sigma$ of Fig.\ \ref{Fig:BasisStc}: The apparently asymptotic
$r_{s}$ is $1.3\sigma$ in Fig.\ \ref{Fig:Bnd12Stc} and $1.7\sigma$
in Fig.\ \ref{Fig:Bnd28Stc}. It is of course natural that the ideal
$r_{s}$ for the short dumbbells is less than the ideal $r_{s}$ for
the long dumbbells, since the former are more ``sphere-like'' than
the latter. This further means that RelRes works better as the molecular
size decreases. As a further analysis, we compute $S_{JS}$ for these
radial distributions, plotting them in Fig.\ \ref{Fig:EntropyDmbl}
as orange (downward) triangles for $\ell=0.3\sigma$ and as green
(upward) triangles for $\ell=0.7\sigma$. For these two cases, we
again observe an essentially exponential decay. We also reaffirm that
the efficacy of RelRes is more deficient for the long molecules than
for the short molecules, considering that $S_{JS}$ for $\ell=0.3\sigma$
is less than that for $\ell=0.7\sigma$ by about an order of magnitude;
remember that $S_{JS}\rightarrow0$ means that we are approaching
perfect replication of the radial distribution. 

In the context of Fig.\ \ref{Fig:BasisStc} for $\ell=0.5\sigma$,
we have thoroughly discussed that the recommended value of $r_{s}$
is signaled by structural correlations. We find that this is still
the case here once we vary the bond length of the dumbbells. Most
importantly, in the bottom panels of Figs.\ \ref{Fig:Bnd12Stc} and
\ref{Fig:Bnd28Stc}, we again notice the flattening of $\left\langle \left(\vec{s}_{i}\cdot\vec{s}_{j}\right)^{2}\right\rangle $
beyond a certain distance, which stems in the mirroring behavior of
the maxima and minima of its components $\left\langle \left(s_{i}^{\Vert}s_{j}^{\Vert}\right)^{2}\right\rangle $
and $\left\langle \left(s_{i}^{\bot}s_{j}^{\bot}\right)^{2}\right\rangle $
that cancel each other. We summarize all the critical distances associated
with these orientational correlations, as well as with the translational
ones in the respective top panels, in Tab.\ \ref{Tab:Distances}.
Among these different bond lengths, we generally notice the same trends:
The inflection in $\left\langle \left(\vec{s}_{i}\cdot\vec{s}_{j}\right)^{2}\right\rangle $
comes just before the inflection in $g$, and this is followed consecutively
by the the minimum in $\left\langle \left(s_{i}^{\bot}s_{j}^{\bot}\right)^{2}\right\rangle $
and the maximum in $\left\langle \left(s_{i}^{\Vert}s_{j}^{\Vert}\right)^{2}\right\rangle $,
with the midpoint between the extrema in $g$ being around there as
well. For a given bond length, the most important aspect of these
distances is that they almost all occur within $0.1\sigma$ of each
other, and they are roughly the same as the estimate for the asymptotic
$r_{s}$ via our systematic RelRes examination. 

As such, it appears that if one is interested in determining the ideal
switching distance, there may be no need of constructing many RelRes
systems with different $r_{s}$: One can just look at the structural
correlations of the reference liquid, with the radial distribution
being perhaps the most feasible choice. As such, we consequently again
draw vertical lines in Figs.\ \ref{Fig:Bnd12Stc} and \ref{Fig:Bnd28Stc}
that represent their inflections in $g$. Nevertheless, we can even
go further: For such dumbbells of an arbitrary bond length, we may
not even need to do any molecular simulations of the reference liquid
at all for determining $r_{s}$ , since it is quite obvious here that
the following linear approximation holds:

\begin{equation}
r_{s}\approx\sigma+\ell\label{Eq:SwitchingDistance}
\end{equation}
This further suggests that for an arbitrary system, one may be able
to do a rough estimate for the ideal $r_{s}$ in RelRes just by considering
the size of the respective molecules.

Finally, we also calculate thermal properties for these two systems.
Analogous with the elementary dumbbells, we plot their average $\left\langle \widetilde{U}\right\rangle $
and variance $\left\langle \delta\widetilde{U}^{2}\right\rangle $
in the bottom and top panels, respectively, of Fig.\ \ref{Fig:EnergyDmbl},
with the orange (downward) triangles for $\ell=0.3\sigma$ and the
green (upward) triangles for $\ell=0.7\sigma$. We again notice the
fluctuating trend in all sets here. Importantly, notice that the magnitude
of these fluctuations decreases with bond length. This reiterates
the fact the RelRes works better for ``sphere-like'' molecules. 

\subsection{\textit{Examining Nonuniform Dumbbells }\textit{\normalsize{}\label{SecB:Computational5}}}

\begin{figure}[H]
\begin{centering}
\includegraphics{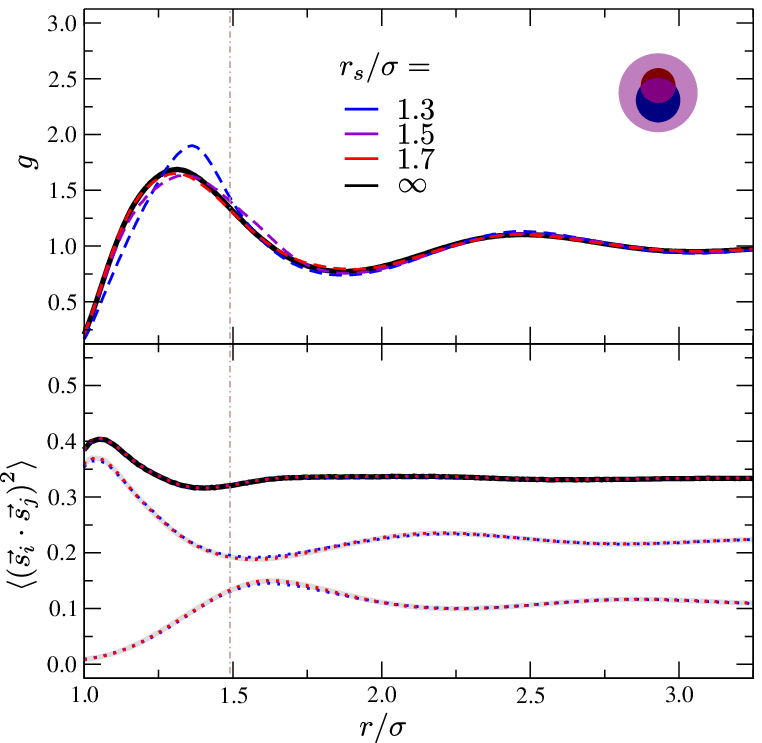}
\par\end{centering}
\caption{\textit{\label{Fig:NufrmStc} }Static correlations for the nonuniform
dumbbells, the ones with different LJ parameters across their two
sites, $\left\{ 0.85\sigma,1.18\sigma\right\} $ and $\left\{ 1.61\epsilon,0.62\epsilon\right\} $.
Everything here is essentially equivalent with Fig.\ \ref{Fig:BasisStc}.
The vertical line is at $1.49$. }
\end{figure}

We now construct another system, whose molecules again have a bond
length of $0.5\sigma$ . In this case however, we are dealing with
nonuniform dumbbells in terms of their LJ parameters, although they
do still have equal mass across their sites; a sketch of such a molecule
is given in Fig.\ \ref{Fig:NufrmStc}. Elaborating on this nonuniformity,
one FG site is small yet strong, and one FG site is large yet weak:
Their respective length parameters are $6/\left(1+\sqrt{37}\right)\approx0.85$
and $\left(1+\sqrt{37}\right)/6\approx1.18$, in units of $\sigma$,
and their respective energy parameters are $\left(1+\sqrt{5}\right)/2\approx1.61$
and $2/\left(1+\sqrt{5}\right)\approx0.62$, in units of $\epsilon$;
notice that the mixed interaction between these two still has the
standard $\sigma$ and $\epsilon$ parameters. Using MPIL, we obtain
that the ideal CG site has LJ parameters that are roughly $1.08\sigma$
and $2.76\epsilon$. Importantly, realize that even though these dumbbells
are nonuniform, the system itself is still uniform, being a single-component
and single-phase liquid. Besides, the density for this system is $0.88\sigma^{3}/m$. 

For this dumbbell system, we again perform a systematic examination
for the switching distance. We present the corresponding structural
correlations in Fig.\ \ref{Fig:NufrmStc}, whose format is identical
with that of Fig.\ \ref{Fig:BasisStc}. We again notice the same
trends that we have observed for our elementary dumbbells. For the
orientational functions of the bottom panel, the replication is perfect
irrespective of the value of $r_{s}$, just as we found in all of
our other scenarios. Most importantly, for the radial distributions
in the top panel, we find that the asymptotic $r_{s}$ is at $1.5\sigma$,
just as we have for our elementary dumbbells (i.e., $\ell=0.5\sigma$).
Once we determine all the critical distances in $g$, as well as in
$\left\langle \left(\vec{s}_{i}\cdot\vec{s}_{j}\right)^{2}\right\rangle $
with its components, we again find that they are all roughly the same
as the ideal $r_{s}$ for RelRes; these are summarized in Tab.\ \ref{Tab:Distances}.
Realize now that Eq.\ \ref{Eq:SwitchingDistance} still applies for
this case, even though we established it while only considering uniform
dumbbells. Finally, once we invoke our functionals for the radial
distributions, we show in Fig.\ \ref{Fig:EntropyDmbl} that for all
$r_{s}$, $S_{JS}$ for this nonuniform scenario (i.e., the almost
hidden yellow diamonds) is basically identical with its uniform counterpart
(i.e., the indigo circles). 

All of these observations for the structural correlations of the nonuniform
dumbbells, particularly the fact that their relationship with the
switching distance of RelRes is identical with that of our elementary
dumbbells (i.e., those with the same bond length of $0.5\sigma$,
yet which are uniform across their sites), has significant ramifications.
For an arbitrary molecule which has much uniformity within it, we
do not need to perform any molecular simulations for it for determining
its ideal $r_{s}$; all one must do is to construct a system of similar
molecules that are actually uniform across their sites (perhaps using
the overall mean values for $\sigma$ and $\epsilon$ of the nonuniform
molecule). According to our observations, these uniform molecules,
whose structural correlations are fairly feasible for computation,
would roughly have the same $r_{s}$ as the nonuniform molecules.
This fact becomes especially useful once we are dealing with a set
of molecules that roughly have the same topology, yet whose sites
notably vary in their LJ parameters: If the molecules do not have
the linear estimate for $r_{s}$ as given in Eq.\ \ref{Eq:SwitchingDistance},
we suggest that all one does is measure, via a single molecular simulation,
the radial distribution of the reference liquid of the uniform version
of these molecules, and we recommend using its inflection in $g$
as the switching distance in RelRes for the entire family of molecules.

We now proceed with the thermal properties of the nonuniform scenario
in Fig.\ \ref{Fig:EnergyDmbl}, which is presented as yellow diamonds.
In this case, we actually do not observe the fluctuating trend which
we have for all other dumbbell scenario, and instead we have a slight
asymptotic behavior; of course, it is very modest as compared with
the entropic functionals of Fig.\ \ref{Fig:EntropyDmbl}. Still,
this is not necessarily a drawback for RelRes: If we look at our recommended
$r_{s}$ of $1.5\sigma$, the variance in the top panel is off by
a negligible fraction, yet the average of the bottom panel is off
by a $\sim0.1$ fraction. Of course, the latter is still satisfactory,
especially if one is not too interested in exact values of thermal
properties. The discrepancy between the ability of RelRes in retrieving
the average and variance may stem in the fact that the latter are
related with response functions, and it is well known that these can
be perfectly captured if the structural correlations are also perfectly
captured, which is our case in this scenario. In fact, the main message
here is that while structural correlations of nonuniform molecules
can be as feasible of retrieving as those of uniform molecules, nonuniform
energetics involve much intricacies with them, and thus a perfect
replication is rather difficult to achieve.

\begin{figure}[H]
\begin{centering}
\includegraphics{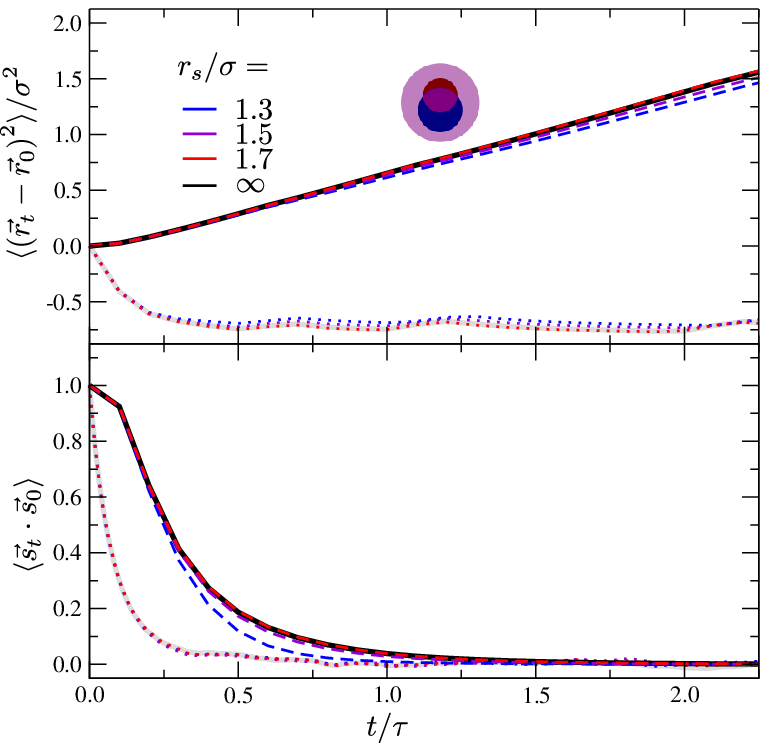}
\par\end{centering}
\caption{\textit{\label{Fig:NufrmDnc} }Dynamic correlations for the nonuniform
dumbbells, the ones with different LJ parameters across their two
sites, $\left\{ 0.85\sigma,1.18\sigma\right\} $ and $\left\{ 1.61\epsilon,0.62\epsilon\right\} $.
Everything here is essentially equivalent with Fig.\ \ref{Fig:BasisDnc}. }
\end{figure}

Finally, we examine the transport functions of this nonuniform scenarios;
we present them in Fig.\ \ref{Fig:NufrmDnc}, which is identical
in format with Fig.\ \ref{Fig:BasisDnc}. We again reaffirm our earlier
observations. Foremost, RelRes perfectly captures the time behavior
of the thermal property of the reference system, regardless of the
switching distance. While the transport functions of the structural
correlations are also adequately retrieved, there are some subtleties:
For perfect replication, $\left\langle \left(\vec{r}_{t}-\vec{r}_{0}\right)^{2}\right\rangle $
of the top panel requires a relatively large $r_{s}$, yet $\left\langle \vec{s}_{t}\cdot\vec{s}_{0}\right\rangle $
of the bottom panel requires a relatively small $r_{s}$. In summary,
just as RelRes is successful in capturing static behavior, it is also
successful in capturing dynamic behavior, and we basically observe
the same capability in describing the nonuniform dumbbells, as with
their uniform counterparts. 

\pagebreak{}
\begin{singlespace}

\section{\textit{Conclusion \label{SecA:Conclusion}}}
\end{singlespace}

In this current work, we continue our initial effort \citep{ChaimovichKremer_JCP2015},
presenting in turn a comprehensive picture of RelRes with MPIL. As
shown in our original communication \citep{ChaimovichKremer_JCP2015},
our hybrid formalism can be thought of as an extension of the common
approach in statistical mechanics which describes all interactions
beyond a certain distance via a mean field \citep{Frisch_ACP1964,Widom_Science1967,WeeksAndersen_JCP1971,IngebrigtsenDyre_PRX2012}.
The main aspect of RelRes is that molecular resolution switches in
terms of relative separation: Each molecule embodies both FG and CG
models, and a given molecule then interacts with near neighbors via
its FG potential and with far neighbors via its CG potential. Via
energy conservation, we relate the FG and CG potentials by an arithmetic
parameterization, which we call MPIL. 

In our current publication, we specifically cast MPIL as a multipole
approximation. Reminiscent of the familiar scenario, we do this derivation
by assuming that the potential can be cast as a linear combination
of inverse powers for the relative separation, and we then perform
the common Taylor expansion. We call the leading nonzero term the
MPIL, and we show that it naturally fits with RelRes. Note that in
our original publication, we naively assumed infinite separation,
in turn readily attaining MPIL \citep{ChaimovichKremer_JCP2015};
this current work is in fact the comprehensive derivation, in which
we specifically present the first and second terms of the Taylor series.
In our original publication, we already showed that RelRes with MPIL
works incredibly well for nonpolar molecules, as complex as a generic
tetrachloromethane-thiophene fluid mixture which contains a vacuum
cavity \citep{ChaimovichKremer_JCP2015}. In the current work, we
take a better examination at the role of the switching distance, finding
an ideal value for it, which can be usually obtained by running a
single molecular simulation of the reference mixture. Besides, we
also show that RelRes can describe not only statics but also dynamics. 

You may notice that RelRes with MPIL bears much resemblance with the
famous ``cell-multipole'' formalism, the algorithm which is frequently
mentioned as one of the most computationally powerful methods \citep{GreengardRokhlin_JCP1987}.
For an arbitrary potential of an inverse power, the ``cell-multipole''
approach also invokes a multipole approximation at appropriate distances:
Inside a ``cell'' at small separations, one evaluates interactions
between discrete sites, and outside a ``cell'' at large separations,
one evaluates interactions between continuum patches. A crucial distinction
with our multiscale scheme is that the ``cell-multipole'' strategy
considers interaction sites that move freely of each other, in turn
lumping them together in arbitrary ``cells'' in space. In molecular
simulations, interaction sites do not move freely of each other as
they are constrained by bonds; RelRes, as well as many other multiscale
methods \citep{PraprotnikKremer_JCP2005,Abrams_JCP2005,EnsingParrinello_JCTC2007,PotestioDonadio_PRL2013},
uses this signature of molecules, invoking a natural mapping from
the FG model to the CG model (i.e., from atomistic coordinates to
gravitational centers, respectively). As such, one can think of RelRes
with MPIL as an natural modification of the ``cell-multipole'' approach
for molecular simulations. This can be extremely useful, since, despite
the fact that the ``cell-multipole'' method is frequently used in
a myriad of systems (e.g., galaxies), it is rarely employed in molecular
systems. 

As a next step, we are interested in implementing RelRes for polar
molecules, specifically for water. Importantly, all the necessary
mathematical ingredients have been already derived here. Unlike the
nonpolar case, our model for water will have an orientational component,
as manifested by its dipole. As we move to biology, we also expect
to model various ``chains''. Of course, RelRes is only one route
for enhancing the efficiency of molecular simulations, and we recommend
that one employs it concurrently with other computational approaches
for optimal results. For biological systems (e.g., a protein in water),
we expect that combining RelRes with explicit-implicit strategies
may be especially powerful: Reminiscent of Adaptive Resolution \citep{WagonerPande_JCP2011},
these algorithms switch from an explicit solute near to the origin
to an implicit solvent far from the origin \citep{BeglovRoux_JCP1994};
the former is the one that is ideal for being modeled by RelRes. In
summary, we emphasize that RelRes with MPIL can be of much use in
aiding the study of soft matter via molecular simulations. 

\pagebreak{}
\begin{singlespace}

\section*{\textit{\normalsize{}Acknowledgments \label{SecA:Acknowledgments}}}
\end{singlespace}

Foremost, we are thankful for partial funding by the Alexander von
Humboldt Foundation, as well as the financial support of the Melvin
J.\ Berman Hebrew Academy. We also appreciate the insightful conversations
with Denis Andrienko regarding polar systems. Finally, we are grateful
for the special services of Mark Wilson. 

\pagebreak{}
\begin{singlespace}

\section*{\textit{\normalsize{}Appendix \label{SecA:Appendix}}}
\end{singlespace}

For proceeding beyond the monopole-monopole term of Eq.\ \ref{Eq:Energy0_},
we must define several dimensionless variables, reminiscent of Eqs.\ \ref{Eq:DmnlsXi}
and \ref{Eq:DmnlsTh}. Here they are:

\begin{equation}
\overrightarrow{\xi}_{\upsilon_{l}}=\frac{\overrightarrow{\Delta}_{\upsilon_{l}}}{r}\label{Eq:DmnlsXi_}
\end{equation}

\begin{equation}
\cos\theta_{\upsilon_{l}}=\frac{\overrightarrow{r}\cdot\overrightarrow{\Delta}_{\upsilon_{l}}}{r\Delta_{\upsilon_{l}}}\label{Eq:DmnlsTh_}
\end{equation}
For compactness in the Appendix, we mostly use here the index $\upsilon$,
together with the index $l$: The Latin index corresponds with either
$i$ or $j$, and the Greek index corresponds with either $\mu$ or
$\nu$. Besides, we also frequently omit yet imply the combination
of the $ij$ indices (e.g., $\overrightarrow{r}=\overrightarrow{r}_{ij}$).
Using these, together with Eqs.\ \ref{Eq:DmnlsXi} and \ref{Eq:DmnlsTh},
we derive useful expressions for the dimensionless variables involved
in the energy function of Eq.\ \ref{Eq:EnergyB}: 

\begin{equation}
\cos\theta_{\mu_{i}\nu_{j}}\xi_{\mu_{i}\nu_{j}}=\cos\theta_{\nu_{j}}\xi_{\nu_{j}}-\cos\theta_{\mu_{i}}\xi_{\mu_{i}}\label{Eq:DmnlsA}
\end{equation}

\begin{equation}
\xi_{\mu_{i}\nu_{j}}^{2}=\xi_{\nu_{j}}^{2}-2\left(\overrightarrow{\xi}_{\nu_{j}}\cdot\overrightarrow{\xi}_{\mu_{i}}\right)+\xi_{\nu_{j}}^{2}\label{Eq:DmnlsB}
\end{equation}
Note that we also used here the definition of $\overrightarrow{\Delta}_{\mu_{i}\nu_{j}}$
of Eq.\ \ref{Eq:Delta}. 

Let us now proceed by invoking these variables in $c_{ij,\aleph}$.
For $\aleph=1$, we specifically substitute Eq.\ \ref{Eq:DmnlsA}
in Eq.\ \ref{Eq:CoefficientTerm1}, attaining the following, 

\begin{equation}
c_{ij,1}\left(\xi_{\mu_{i}\nu_{j}},\cos\theta_{\mu_{i}\nu_{j}}\right)=m\sum_{\mu_{i}\nu_{j}}\left[c_{\mu_{i}\nu_{j}}\cos\theta_{\nu_{j}}\xi_{\nu_{j}}\right]-m\sum_{\mu_{i}\nu_{j}}\left[c_{\mu_{i}\nu_{j}}\cos\theta_{\mu_{i}}\xi_{\mu_{i}}\right]\label{Eq:CoefficientTerm1A}
\end{equation}
and by invoking the product assumption of Eq.\ \ref{Eq:Coefficient},
we derive this: 

\begin{equation}
c_{ij,1}=m\left(\sum_{\mu_{i}}\mathfrak{c}_{\mu_{i}}\right)\left(\sum_{\nu_{j}}\left[\mathfrak{c}_{\nu_{j}}\cos\theta_{\nu_{j}}\xi_{\nu_{j}}\right]\right)-m\left(\sum_{\nu_{j}}\mathfrak{c}_{\nu_{j}}\right)\left(\sum_{\mu_{i}}\left[\mathfrak{c}_{\mu_{i}}\cos\theta_{\mu_{i}}\xi_{\mu_{i}}\right]\right)\label{Eq:CoefficientTerm1B}
\end{equation}
For facilitating the ensuing mathematics, we introduce now $c_{ij,2\mathfrak{p}^{\pm}}$
and $c_{ij,2\mathfrak{q}^{\pm}}$, whose summation yields $c_{ij,2^{\pm}}$:

\begin{equation}
c_{ij,2^{\pm}}=c_{ij,2\mathfrak{p}^{\pm}}+c_{ij,2\mathfrak{q}^{\pm}}\label{Eq:CoefficientTerm2__}
\end{equation}
For $\aleph=2^{+}$, we specifically substitute Eq.\ \ref{Eq:DmnlsA}
in Eq.\ \ref{Eq:CoefficientTerm2+}, attaining the following, 

\begin{equation}
c_{ij,2\mathfrak{p}^{+}}\left(\xi_{\mu_{i}\nu_{j}},\cos\theta_{\mu_{i}\nu_{j}}\right)=-m\sum_{\mu_{i}\nu_{j}}\left[c_{\mu_{i}\nu_{j}}\cos\theta_{\nu_{j}}\xi_{\nu_{j}}\cos\theta_{\mu_{i}}\xi_{\mu_{i}}\right]\label{Eq:CoefficientTerm2Ap+}
\end{equation}

\begin{equation}
c_{ij,2\mathfrak{q}^{+}}\left(\xi_{\mu_{i}\nu_{j}},\cos\theta_{\mu_{i}\nu_{j}}\right)=\frac{m}{2}\sum_{\mu_{i}\nu_{j}}\left[c_{\mu_{i}\nu_{j}}\cos^{2}\theta_{\nu_{j}}\xi_{\nu_{j}}^{2}\right]+\frac{m}{2}\sum_{\mu_{i}\nu_{j}}\left[c_{\mu_{i}\nu_{j}}\cos^{2}\theta_{\mu_{i}}\xi_{\mu_{i}}^{2}\right]\label{Eq:CoefficientTerm2Aq+}
\end{equation}
and by invoking the product assumption of Eq.\ \ref{Eq:Coefficient},
we derive this: 

\begin{equation}
c_{ij,2\mathfrak{p}^{+}}=-m\left(\sum_{\nu_{j}}\left[\mathfrak{c}_{\nu_{j}}\cos\theta_{\nu_{j}}\xi_{\nu_{j}}\right]\right)\left(\sum_{\mu_{i}}\left[\mathfrak{c}_{\mu_{i}}\cos\theta_{\mu_{i}}\xi_{\mu_{i}}\right]\right)\label{Eq:CoefficientTerm2Bp+}
\end{equation}

\begin{equation}
c_{ij,2\mathfrak{q}^{+}}=\frac{m}{2}\left(\sum_{\mu_{i}}\mathfrak{c}_{\mu_{i}}\right)\left(\sum_{\nu_{j}}\left[\mathfrak{c}_{\nu_{j}}\cos^{2}\theta_{\nu_{j}}\xi_{\nu_{j}}^{2}\right]\right)+\frac{m}{2}\left(\sum_{\nu_{j}}\mathfrak{c}_{\nu_{j}}\right)\left(\sum_{\mu_{i}}\left[\mathfrak{c}_{\mu_{i}}\cos^{2}\theta_{\mu_{i}}\xi_{\mu_{i}}^{2}\right]\right)\label{Eq:CoefficientTerm2Bq+}
\end{equation}
For $\aleph=2^{-}$, we specifically substitute Eq.\ \ref{Eq:DmnlsB}
in Eq.\ \ref{Eq:CoefficientTerm2-}, attaining the following, 

\begin{equation}
c_{ij,2\mathfrak{p}^{-}}\left(\xi_{\mu_{i}\nu_{j}},\cos\theta_{\mu_{i}\nu_{j}}\right)=-m\sum_{\mu_{i}\nu_{j}}\left[c_{\mu_{i}\nu_{j}}\left(\overrightarrow{\xi}_{\nu_{j}}\cdot\overrightarrow{\xi}_{\mu_{i}}\right)\right]\label{Eq:CoefficientTerm2Ap-}
\end{equation}

\begin{equation}
c_{ij,2\mathfrak{q}^{-}}\left(\xi_{\mu_{i}\nu_{j}},\cos\theta_{\mu_{i}\nu_{j}}\right)=\frac{m}{2}\sum_{\mu_{i}\nu_{j}}\left[c_{\mu_{i}\nu_{j}}\xi_{\nu_{j}}^{2}\right]+\frac{m}{2}\sum_{\mu_{i}\nu_{j}}\left[c_{\mu_{i}\nu_{j}}\xi_{\mu_{i}}^{2}\right]\label{Eq:CoefficientTerm2Aq-}
\end{equation}
and by invoking the product assumption of Eq.\ \ref{Eq:Coefficient},
we derive this: 

\begin{equation}
c_{ij,2\mathfrak{p}^{-}}=-m\left(\left(\sum_{\nu_{j}}\left[\mathfrak{c}_{\nu_{j}}\overrightarrow{\xi}_{\nu_{j}}\right]\right)\cdot\left(\sum_{\mu_{i}}\left[\mathfrak{c}_{\mu_{i}}\overrightarrow{\xi}_{\mu_{i}}\right]\right)\right)\label{Eq:CoefficientTerm2Bp-}
\end{equation}

\begin{equation}
c_{ij,2\mathfrak{q}^{-}}=\frac{m}{2}\left(\sum_{\mu_{i}}\mathfrak{c}_{\mu_{i}}\right)\left(\sum_{\nu_{j}}\left[\mathfrak{c}_{\nu_{j}}\xi_{\nu_{j}}^{2}\right]\right)+\frac{m}{2}\left(\sum_{\nu_{j}}\mathfrak{c}_{\nu_{j}}\right)\left(\sum_{\mu_{i}}\left[\mathfrak{c}_{\mu_{i}}\xi_{\mu_{i}}^{2}\right]\right)\label{Eq:CoefficientTerm2Bq-}
\end{equation}
While the monopoles of Eq.\ \ref{Eq:Monopole} can be readily substituted
in most of the above expressions, further manipulation must be performed
for employing here the dipoles, as well as the quadrupoles. 

As such, let us evaluate some of the summations which appear above,
by invoking the dimensionless variables of Eqs.\ \ref{Eq:DmnlsXi_}
and \ref{Eq:DmnlsTh_}. Here are summations which can be conveniently
cast in terms of the dipoles of Eq.\ \ref{Eq:Dipole},

\begin{equation}
\sum_{\upsilon_{l}}\left[\mathfrak{c}_{\upsilon_{l}}\cos\theta_{\upsilon_{l}}\xi_{\upsilon_{l}}\right]=\frac{1}{r^{2}}\sum_{\upsilon_{l}}\left[\mathfrak{c}_{\upsilon_{l}}\left(\overrightarrow{r}\cdot\overrightarrow{\Delta}_{\upsilon_{l}}\right)\right]=\frac{1}{r^{2}}\left(\overrightarrow{r}\cdot\left(\sum_{\upsilon_{l}}\left[\mathfrak{c}_{\upsilon_{l}}\overrightarrow{\Delta}_{\upsilon_{l}}\right]\right)\right)=\frac{1}{r}\left(\hat{r}\cdot\mathfrak{\overrightarrow{p}}_{l,1}\right)\label{Eq:SumA}
\end{equation}

\begin{equation}
\sum_{\upsilon_{l}}\left[\mathfrak{c}_{\upsilon_{l}}\overrightarrow{\xi}_{\upsilon_{l}}\right]=\frac{1}{r}\sum_{\upsilon_{l}}\left[\mathfrak{c}_{\upsilon_{l}}\overrightarrow{\Delta}_{\upsilon_{l}}\right]=\frac{1}{r}\mathfrak{\overrightarrow{p}}_{l,1}\label{Eq:SumB}
\end{equation}
while here are summations which can be conveniently cast in terms
of the quadrupoles of Eqs.\ \ref{Eq:Quadrupole+} and \ref{Eq:Quadrupole-}; 

\begin{equation}
\sum_{\upsilon_{l}}\left[\mathfrak{c}_{\upsilon_{l}}\cos^{2}\theta_{\upsilon_{l}}\xi_{\upsilon_{l}}^{2}\right]=\frac{1}{r^{4}}\sum_{\upsilon_{l}}\left[\mathfrak{c}_{\upsilon_{l}}\left(\overrightarrow{r}\cdot\overrightarrow{\Delta}_{\upsilon_{l}}\right)^{2}\right]=\frac{1}{r^{4}}\left(\overrightarrow{r}\overrightarrow{r}:\left(\sum_{\upsilon_{l}}\left[\mathfrak{c}_{\upsilon_{l}}\overrightarrow{\Delta}_{\upsilon_{l}}\overrightarrow{\Delta}_{\upsilon_{l}}\right]\right)\right)=\frac{1}{r^{2}}\left(\hat{\hat{r}}:\mathfrak{\overrightarrow{\overrightarrow{q}}}_{l,2}\right)\label{Eq:SumC}
\end{equation}

\begin{equation}
\sum_{\upsilon_{l}}\left[\mathfrak{c}_{\upsilon_{l}}\xi_{\upsilon_{l}}^{2}\right]=\frac{1}{r^{2}}\sum_{\upsilon_{l}}\left[\mathfrak{c}_{\upsilon_{l}}\Delta_{\upsilon_{l}}^{2}\right]=\frac{1}{r^{2}}\mathfrak{q}_{l,0}\label{Eq:SumD}
\end{equation}
note that we used here a common tensor identity: $\left(\overrightarrow{r}\cdot\overrightarrow{\Delta}_{\upsilon_{l}}\right)^{2}=\left(\overrightarrow{r}\overrightarrow{r}:\overrightarrow{\Delta}_{\upsilon_{l}}\overrightarrow{\Delta}_{\upsilon_{l}}\right)$. 

By employing these, together with Eq.\ \ref{Eq:Monopole}, we can
evaluate the relevant $c_{ij,\aleph}$. For $\aleph=1$, substituting
Eq.\ \ref{Eq:SumA} in Eq.\ \ref{Eq:CoefficientTerm1B}, we obtain
the following: 

\begin{equation}
c_{ij,1}=\frac{m}{r}\mathfrak{c}_{i,0}\left(\hat{r}\cdot\mathfrak{\overrightarrow{p}}_{j,1}\right)-\frac{m}{r}\mathfrak{c}_{j,0}\left(\hat{r}\cdot\mathfrak{\overrightarrow{p}}_{i,1}\right)\label{Eq:CoefficientTerm1__}
\end{equation}
This is compactly presented in the main text as Eq.\ \ref{Eq:CoefficientTerm1_}.
For $\aleph=2^{+}$, substituting Eqs.\ \ref{Eq:SumA} and \ref{Eq:SumC}
in Eqs.\ \ref{Eq:CoefficientTerm2Bp+} and \ref{Eq:CoefficientTerm2Bq+},
respectively, we obtain the following: 

\begin{equation}
c_{ij,2\mathfrak{p}^{+}}=-\frac{m}{r^{2}}\left(\hat{r}\cdot\mathfrak{\overrightarrow{p}}_{j,1}\right)\left(\hat{r}\cdot\mathfrak{\overrightarrow{p}}_{i,1}\right)\label{Eq:CoefficientTerm2p+_}
\end{equation}

\begin{equation}
c_{ij,2\mathfrak{q}^{+}}=\frac{m}{2r^{2}}\mathfrak{c}_{i,0}\left(\hat{\hat{r}}:\mathfrak{\overrightarrow{\overrightarrow{q}}}_{j,2}\right)+\frac{m}{2r^{2}}\mathfrak{c}_{j,0}\left(\hat{\hat{r}}:\mathfrak{\overrightarrow{\overrightarrow{q}}}_{i,2}\right)\label{Eq:CoefficientTerm2q+_}
\end{equation}
For $\aleph=2^{-}$, substituting Eqs.\ \ref{Eq:SumB} and \ref{Eq:SumD}
in Eqs.\ \ref{Eq:CoefficientTerm2Bp-} and \ref{Eq:CoefficientTerm2Bq-},
respectively, we obtain the following: 

\begin{equation}
c_{ij,2\mathfrak{p}^{-}}=-\frac{m}{r^{2}}\left(\mathfrak{\overrightarrow{p}}_{j,1}\cdot\mathfrak{\overrightarrow{p}}_{i,1}\right)\label{Eq:CoefficientTerm2p-_}
\end{equation}

\begin{equation}
c_{ij,2\mathfrak{q}^{-}}=\frac{m}{2r^{2}}\mathfrak{c}_{i,0}\mathfrak{q}_{j,0}+\frac{m}{2r^{2}}\mathfrak{c}_{j,0}\mathfrak{q}_{i,0}\label{Eq:CoefficientTerm2q-_}
\end{equation}
Invoking Eq.\ \ref{Eq:CoefficientTerm2__}, these are compactly presented
in the main text as Eqs.\ \ref{Eq:CoefficientTerm2+_} and \ref{Eq:CoefficientTerm2-_}.
Besides, in an analogous manner which we performed in this Appendix
for the first and second terms of $c_{ij,\aleph}$, other terms of
this coefficient can be evaluated as well. 

\pagebreak{}

\textcompwordmark{}\\

\pagebreak{}

\begin{table}
\begin{centering}
\begin{tabular}{c>{\centering}m{36mm}>{\centering}m{24mm}>{\centering}m{24mm}>{\centering}m{24mm}>{\centering}m{24mm}}
\toprule 
 & Molecule Type & Elementary & Short & Long & Nonuniform\tabularnewline
\midrule 
\multirow{1}{*}{} & Bond Length & $0.5$ & $0.3$ & $0.7$ & $0.5$\tabularnewline
\midrule
\midrule 
 & Suggested $r_{s}$ & $1.5$ & $1.3$ & $1.7$ & $1.5$\tabularnewline
\midrule
\midrule 
 & Inflection in $g$ & $1.46$ & $1.28$ & $1.64$ & $1.49$\tabularnewline
\midrule 
\multirow{1}{*}{} & Midpoint between

extrema in $g$ & $1.54$ & $1.43$ & $1.66$ & $1.59$\tabularnewline
\midrule
\midrule 
 & Inflection in 

$\left\langle \left(\vec{s}_{i}\cdot\vec{s}_{j}\right)^{2}\right\rangle $  & $1.42$ & $1.25$ & $1.59$ & $1.51$\tabularnewline
\midrule 
 & Maximum in 

$\left\langle \left(s_{i}^{\Vert}s_{j}^{\Vert}\right)^{2}\right\rangle $  & $1.49$ & $1.36$ & $1.64$ & $1.57$\tabularnewline
\midrule 
 & Minimum in 

$\left\langle \left(s_{i}^{\bot}s_{j}^{\bot}\right)^{2}\right\rangle $ & $1.54$ & $1.39$ & $1.69$ & $1.65$\tabularnewline
\bottomrule
\end{tabular}
\par\end{centering}
\caption{\label{Tab:Distances} Signature distances in the structural correlations
of the various dumbbell scenarios. Rows 1-2 give us the molecule type,
as well as its respective bond length, which each column corresponds
with. Column 2 goes together with Fig.\ \ref{Fig:BasisStc}, columns
3 and 4 refer to Figs.\ \ref{Fig:Bnd12Stc} and \ref{Fig:Bnd28Stc},
respectively, and column 5 goes with Fig.\ \ref{Fig:NufrmStc}. Column
1 then tells us which distance value we are dealing with in each case.
Row 3 is the switching distance that we recommend for use in RelRes,
based on our systematic examination of varying $r_{s}$. Rows 4-5
deal with signatures of the radial distributions, while rows 6-8 deal
with signatures of the orientational functions. Realize that all numbers
here are given in units of $\sigma$.}
\end{table}

\textcompwordmark{}\\

\pagebreak{}
\begin{singlespace}

\section*{\textit{\normalsize{}Bibliography \label{SecA:Bibliography}}}
\end{singlespace}

\begin{singlespace}
\renewcommand{\section}[2]{}

{\small{}\bibliographystyle{unsrt}
\bibliography{Manuscript}
}{\small \par}
\end{singlespace}

\end{document}